\documentclass[aps,prr, twocolumn, floatfix,superscriptaddress]{revtex4-2}

\pdfoutput = 1
\usepackage{graphicx}
\usepackage{hyperref}
\hypersetup{hidelinks}
\usepackage{amsmath}
\usepackage{multirow}
\usepackage{amssymb}
\usepackage{bbm}
\usepackage{color}
\usepackage{dcolumn}
\usepackage{bm}
\usepackage{float}
\usepackage[normalem]{ulem}
\usepackage{braket}
\usepackage{changepage}
\usepackage[subrefformat=parens,labelformat=parens,caption=false]{subfig}
\usepackage{bbm}
\usepackage[dvipsnames]{xcolor}
\usepackage{esint}
\usepackage{lineno}
\usepackage{verbatim}
\usepackage{bm}
\usepackage{bbold}
\usepackage{natbib}

\DeclareMathOperator{\Tr}{Tr}

\usepackage[subrefformat=parens,labelformat=parens,caption=false]{subfig}

\usepackage[export]{adjustbox}

\newcommand{\jani}[1]%
   {\begingroup{\color{Mahogany}Jani: \textit{#1}}\endgroup}
   
\newcommand{\manish}[1]%
   {\begingroup{\color{Lavender}Manish: \textit{#1}}\endgroup}
\usepackage{soul}

\usepackage{tikz,xcolor,hyperref}

\definecolor{lime}{HTML}{A6CE39}

\begin{document}
\title{Co-Design quantum simulation of nanoscale NMR}

\author{Manuel G. Algaba}
\thanks{Both authors contributed equally to this work. \\ Corresponding author: manuel.algaba@meetiqm.com \\ Corresponding author: mario.ponce@meetiqm.com}
\affiliation{IQM Quantum Computers, Nymphenburgerstr. 86, 80636 Munich, Germany}
\author{Mario Ponce-Martinez}
\thanks{Both authors contributed equally to this work. \\ Corresponding author: manuel.algaba@meetiqm.com \\ Corresponding author: mario.ponce@meetiqm.com}
\affiliation{IQM Quantum Computers, Nymphenburgerstr. 86, 80636 Munich, Germany}
\affiliation{Department of Physics and Arnold Sommerfeld Center for Theoretical Physics, Ludwig-Maximilians-Universit\" at M\" unchen, Theresienstrasse 37, 80333 Munich, Germany}
\author{Carlos Munuera-Javaloy}
\affiliation{Department of Physical Chemistry, University of the Basque Country UPV/EHU, Apartado 644, 48080 Bilbao, Spain}
\author{Vicente Pina-Canelles}
\affiliation{IQM Quantum Computers, Nymphenburgerstr. 86, 80636 Munich, Germany}
\author{Manish J. Thapa}
\affiliation{IQM Quantum Computers, Nymphenburgerstr. 86, 80636 Munich, Germany}
\author{Bruno G. Taketani}
\affiliation{IQM Quantum Computers, Nymphenburgerstr. 86, 80636 Munich, Germany}
\author{Martin Leib}
\affiliation{IQM Quantum Computers, Nymphenburgerstr. 86, 80636 Munich, Germany}
\author{In\'es de Vega}
\affiliation{IQM Quantum Computers, Nymphenburgerstr. 86, 80636 Munich, Germany}
\affiliation{Department of Physics and Arnold Sommerfeld Center for Theoretical Physics, Ludwig-Maximilians-Universit\" at M\" unchen, Theresienstrasse 37, 80333 Munich, Germany}
\author{Jorge Casanova}
\affiliation{Department of Physical Chemistry, University of the Basque Country UPV/EHU, Apartado 644, 48080 Bilbao, Spain}
\affiliation{IKERBASQUE, Basque Foundation for Science, Plaza Euskadi 5, 48009 Bilbao, Spain}
\author{Hermanni Heimonen}
\affiliation{IQM Quantum Computers, Keilaranta 19, FI-02150 Espoo, Finland}

\begin{abstract}

Quantum computers have the potential to efficiently simulate the dynamics of nanoscale NMR systems.
In this work we demonstrate that a noisy intermediate-scale quantum computer can be used to simulate and predict nanoscale NMR resonances. In order to minimize the required gate fidelities, we propose a superconducting application-specific Co-Design quantum processor that 
reduces the number of SWAP gates by over 90\% for chips with more than 20 qubits. The processor consists of transmon qubits capacitively coupled via tunable couplers to a central co-planar waveguide resonator with a quantum circuit refrigerator (QCR) for fast resonator reset. The QCR implements the non-unitary quantum operations required to simulate nuclear hyperpolarization scenarios.

\end{abstract}

\date{\today}
\maketitle

\section{Introduction}

Computer simulations are the backbone of scientific research and technological development. Quantum computers promise in the long term to enable simulations of systems that are intractable to even the largest supercomputers~\cite{feynman2018simulating,lloyd1996universal}. Currently, scientists have access to so-called noisy intermediate-scale quantum (NISQ) computers~\cite{preskill2018quantum}, that present limited qubit counts without error correction. While applications of error-corrected quantum computers are well established, use cases where NISQ devices might achieve quantum advantage are still elusive~\cite{bharti2021noisy}. In the search for these early applications, the problem must fit the hardware, and the hardware must enable implementation with minimal overheads. 

Application-Specific Integrated Chips (ASICs) are highly specialized processors optimized for specific problems when execution speed, power efficiency, or miniaturization is of utmost importance~\cite{smith1997application}. A prominent example where computational speed and energy efficiency are optimised through the use of ASICs is training of artifical neural networks using tensor processing units~\cite{hsu2021gptpu,lu2020accelerating}.
Building a general-purpose quantum computer capable of rivaling the most powerful classical computers has proven to be a difficult task, so it is likely that the first devices reaching useful quantum advantage
will use quantum ASICs, also called Co-Design quantum computers.

A good example of a problem with suitable structure for simulation by quantum computers is nanoscale nuclear magnetic resonance (NMR)~\cite{Staudacher2013}. The problem can be described by a number of mutually interacting spins, which natively map to the qubits of a quantum computer, thereby circumventing the overheads in mapping the problem to qubits, such as in the case of fermions~\cite{nielsen2002quantum}. 
 
In general, fast and reliable quantum simulations of interacting spin systems would improve the interpretability of solid-state NMR and electron spin resonance (ESR) spectra, where advanced numerical techniques present very limited performance~\cite{kuprovspinach}. This shows the potential of quantum computers with a moderate number of qubits to shed light on the dynamics of these important systems. A Co-Design quantum computer that minimizes algorithm implementation overheads could be the first method to access these simulations. Note that, other NMR problems, such as zero-field NMR~\cite{seetharam2021digital} and Hamiltonian learning~\cite{o2021quantum}, have already attracted research on how quantum computers can be used to tackle them and methods based on Bayesian computation~\cite{sels2020quantum} and generative models~\cite{sels2021quantum} have been developed for computing NMR spectra as well.  

NMR techniques have a profound impact in research areas such as material science, chemistry, biology, and medicine~\cite{levitt2013spin}. Recently they have approached the nanoscale through solid-state quantum sensors such as the nitrogen vacancy (NV) center in diamond~\cite{doherty2013nitrogen}. This is a particularly powerful quantum device, as it enables detection and control of nearby nuclear spins with nanoscale resolution~\cite{abobeih2019atomic}. 
Applications of the device are, e.g., the precise determination of the structure and dynamics of nuclear ensembles such as proteins~\cite{MunueraJavaloy2021}, finding inter-label distances (via, e.g., Bayesian analysis of the NV response) in electronically labelled biomolecules~\cite{munuerajavaloy2021detection},
and the exploration of bespoke microwave (MW) sequences that efficiently transfer NV center polarization to the nuclear environment. Hyperpolarization (i.e. polarization beyond that of a thermal state in a magnetic field) of nuclear spins in diamond presents the potential to develop new and safer contrast agents for magnetic resonance imaging. This problem, which we aim to address through simulation by a quantum computer, could lead to improved detection of different malformations in tissues --such as heart or brain-- without the need to deliver ionizing radiation, in contrast to other techniques~\cite{ajoy2018orientation}.

This manuscript describes a Co-Design process for a quantum chip able to efficiently simulate nanoscale NMR scenarios. It is structured in three main parts, each of which is a crucial step in the Co-Design process: 
1. Identifying the problem (Sec.~\ref{sec:hyperpolarization}), which here is simulating a nanoscale NMR system for hyperpolarizing nuclear spins. 2. Choosing an algorithm for the nanoscale NMR problem and showing that a star-topology chip implements it with minimal overhead (Sec.~\ref{sec:algorithm}), and 3. Co-Designing the corresponding quantum chip using a central resonator bus (Sec.~\ref{sec:co-design_HW}). The sections are followed by results and discussions (Sec.~\ref{section:results}) and an outlook (Sec.~\ref{sec:conclusions}).

\section{Nanoscale NMR: Hyperpolarization }
\label{sec:hyperpolarization}

\begin{figure}
    \centering
    \includegraphics[width=0.99\linewidth]{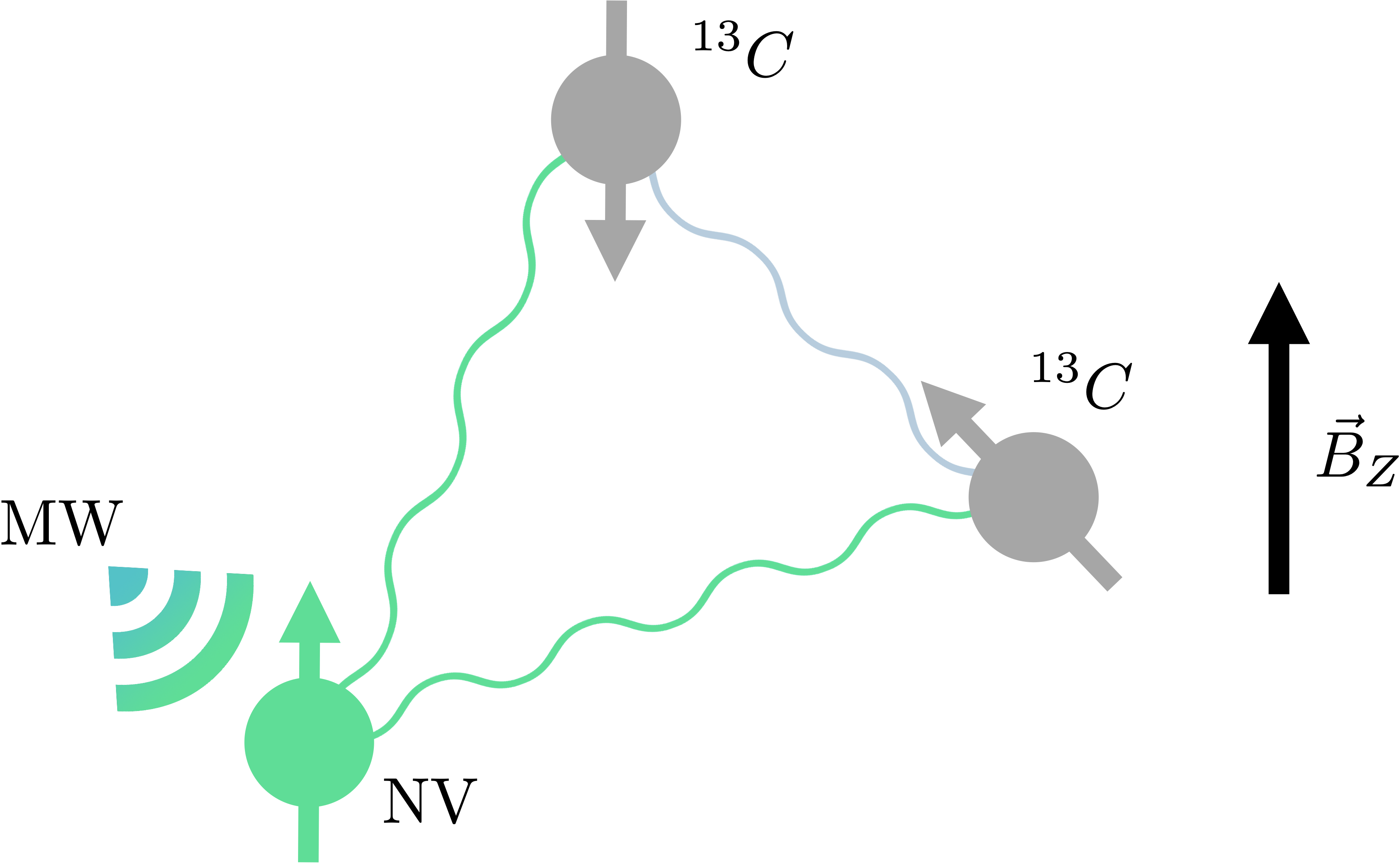}
    \caption{NV center with a microwave drive interacting with two mutually interacting $^{13}C$ nuclei in a magnetic field $\vec{B}_Z$, corresponding to the Hamiltonian in Eq.~\eqref{eq:hamiltonian1} for $M=1$ and $N=2$.}
    \label{fig:interaction_scheme}
\end{figure}

Let us consider a system consisting of $M$ nitrogen-vacancy (NV) centers and $N$ carbon-13 isotopes in the presence of a driving field and an external magnetic field $\vec{B}_Z$. NV centers and nuclei are all effectively described as spin-1/2 systems. The representation of such a system for $M=1$, $N=2$ is shown in Fig.~\ref{fig:interaction_scheme}. For simplicity, we consider the NV centers aligned with the external magnetic field, leading to the following Hamiltonian:

\begin{equation}{\label{eq:hamiltonian1}}
\begin{split}
H = &\sum_{j = 1}^M \delta_j \sigma^z_j-\sum_{k = 1}^N\vec{\omega}^c_k\cdot\vec{I}_k+ \sum_{j = 1}^M\sum_{k = 1}^N\frac{\sigma^z_j}{2} \vec{A}_{jk}\cdot\vec{I}_k+\\&+\sum_{k>k'}^N g_{k'k}\left[I_{k'}^z I_k^z-\frac{1}{4}(I_{k'}^+I_k^-+I_{k'}^-I_k^+)\right]+\\&+\sum_{j>j'}^M h_{j'j}\left[\sigma_{j'}^z \sigma_j^z-2(\sigma_{j'}^+\sigma_j^-+\sigma_{j'}^-\sigma_j^+)\right] + H_{\textrm{dr}}. 
\end{split}
\end{equation}

In Eq.~(\ref{eq:hamiltonian1}) we find the spin operators in the joint Hilbert space $\mathbb{C}^{2^{(M+N)}}$ of NV centers and nuclei: $$\sigma_j^{\mu}=\underbrace{\mathbb{1} \otimes \cdots \otimes \mathbb{1} \otimes \overbrace{\sigma_{\mu}}^{j^{\textrm{th}} \textrm{pos.}} \otimes \, \mathbb{1} \otimes \cdots \otimes \mathbb{1}}_{M \text { factors }} \otimes \underbrace{\mathbb{1} \otimes \ldots \otimes \mathbb{1}}_{N \text { factors }},$$ $$I_k^\mu=\underbrace{\mathbb{1} \otimes \ldots \otimes \mathbb{1}}_{M \text { factors }} \otimes \underbrace{\mathbb{1} \otimes \cdots \otimes \mathbb{1} \otimes \overbrace{\tfrac{1}{2} \sigma_\mu}^{(M+k)^{\textrm{th}} \textrm{pos.}} \otimes \, \mathbb{1} \otimes \cdots \otimes \mathbb{1}}_{N \text { factors }},$$ where $(\sigma_{\mu})_{2\times 2}$, $\mu \in \{x,y,z \}$ is the corresponding $2\times 2$ Pauli matrix on the $j^{\textrm{th}}$ NV center and the $k^{\textrm{th}}$ nucleus respectively, and $\mathbb{1}$ is the $2\times 2$ identity matrix. Accordingly, $\sigma^\pm_j = \frac{\sigma^x_j\pm i\sigma^y_j}{2}\left(I^\pm_k = I^x_k\pm iI^y_k\right)$ are the $j^{\textrm{th}}$ NV center ($k^{\textrm{th}}$ nucleus) ladder operators. The term $\delta_j$ is the detuning of the $j^{_{\textrm{th}}}$ NV center with respect to the microwave drive $H_{\textrm{dr}}$. The hyperfine coupling vector $\vec{A}_{jk}$ represents the coupling between the $j^{_{\textrm{th}}}$ NV center and the $k^{_{\textrm{th}}}$ nucleus, while $\vec{\omega}_k^c = \gamma_c \vec{B}_Z-\frac{1}{2}\sum_{j = 1}^M \vec{A}_{jk}$ is the modified Larmor frequency of the $k^{_{\textrm{th}}}$ nucleus with the $^{13}C$ gyromagnetic ratio $\gamma_c\approx (2\pi) \times 10.7$ MHz/T, $g_{k'k}$ is the coupling between the $k^{_{\textrm{th}}}$ and $k'^{_{\textrm{th}}}$ nuclei, and $h_{j'j}$ is the coupling between the $j^{_{\textrm{th}}}$ and $j'^{_{\textrm{th}}}$ NV centers. 

Note that, Eq.~(\ref{eq:hamiltonian1}) is expressed in a rotating frame with respect to the free NV Hamiltonian, while $H_{\rm dr}$ represents an external driving tuned near resonance with a certain NV energy transition. The derivation of Eq.~(\ref{eq:hamiltonian1}) can be found in Appendix~\ref{appendix:H_derivation}.

In order to hyperpolarize a diamond sample at room temperature, the NV centers are first optically polarized employing laser light, and then their state is transferred to the surrounding nuclei with the aid of a tailored microwave radiation scheme. The initial state of the nuclei in a room-temperature sample is well described by a fully-mixed state due to the small energy splitting of the nuclear spins. By re-initializing the NV centers and repeating this procedure, the polarization transferred into the sample can be amplified. In this paper we will consider the quantum simulation of the polarization transfer mechanism and study two different driving schemes acting on the NV centers in a room-temperature diamond.

The first driving scheme is a continuous driving whose Hamiltonian in the rotating frame mentioned earlier is $H_{\textrm{dr}} =\frac{\Omega}{2} \sigma^{\phi}$, where $\sigma^\phi=e^{-i\phi} |1\rangle\langle 0|+e^{i\phi} |0\rangle\langle 1| = e^{-i\phi}\sigma^{-}+e^{i\phi} \sigma^{+}$, $\phi$ a phase, $\Omega$ the  Rabi frequency and the kets $|1\rangle$ and $|0\rangle$ are the eigenvectors of the operator $\sigma_z$ with eigenvalues $\pm1$ respectively. The set $\{|0\rangle,|1\rangle\}$ is called the computational basis of the state space of a two level system, and will be our standard choice for a basis, $|0\rangle \equiv (1,0)^t$ and $|1\rangle \equiv (0,1)^t$. NV-nucleus polarization transfer is achieved when the Rabi frequency matches the modified nuclear Larmor frequency (i.e. when $ \Omega = |\vec{\omega}_c|$), leading to the Hartmann-Hahn double-resonance condition~\cite{hartmann1962nuclear}.
For a single NV center and nucleus, the Hamiltonian in Eq.~\eqref{eq:hamiltonian1} reduces, in an interaction picture, to $H_I= \frac{A^\perp}{4}\left(|+\rangle\langle-|I^++|-\rangle\langle+|I^-\right)$, where $\ket{\pm} = \ket{0}\pm\ket{1}$, which shows a polarization transfer mechanism with the effective transfer rate $\frac{A^\perp}{4}$ (a detailed derivation can be found in Appendix~\ref{HH_sequence}).

The second type of driving we consider is a pulsed-driving scheme, $H_{\textrm{dr}}=\frac{\Omega(t)}{2}\sigma^\phi$, where $\Omega(t)$ is a train of $\pi$-pulses, such as the Carr-Purcell-Meiboom-Gill sequence~\cite{carr1954effects,meiboom1958modified} or the XY8 sequence~\cite{maudsley1986modified,gullion1990new}. We consider pulses with a negligible width compared to the time spacing $\tau$ between the $\pi$-pulses. If $\tau$ is selected such that $\tau = \frac{n\pi}{|\vec{\omega}^c|}$ ($n$ being an arbitrary integer number) and the pulses are evenly spaced one finds that, in an interaction picture, for a single nucleus and NV center, the Hamiltonian reduces to $H_I = \alpha A^\perp \sigma_z I_x$, where $\alpha$ is a factor that depends on the integer $n$ (see Appendix~\ref{HH_sequence}).  A phase imprinted on the pulse sequence through a time delay turns the interaction into $H_I = \alpha A^\perp \sigma_z I_y$. By combining both sequences with the appropriate rotations over the NV center, the polarization transfer interaction $H_I = -\frac{\alpha A^\perp}{4}\left(\sigma^+I^-+\sigma^-I^+\right)$ is achieved (see Appendix~\ref{HH_sequence} and Ref.~\cite{casanova2016noise} for more details). 

Regarding common error sources, NV centers located at different positions in the diamond lattice experience stress conditions that lead to local energy deviations from the zero-field splitting. The corresponding term in Eq.~\eqref{eq:hamiltonian1} is the detuning $\delta_{j}$. 
Another common type of imperfection appears due to unavoidable fluctuations of the Rabi frequency of the driving. This fluctuation can be modelled as an Ornstein-Uhlenbeck (OU) process~\cite{uhlenbeck1930theory}, which has been shown to be an accurate description for NV centers~\cite{cai2012robust}. It is a Gaussian process of the following form~\cite{OUformula}:

\begin{equation}
\label{eq:OU}
X(t+\Delta t)=X(t) \, \mathrm{e}^{-\Delta t / \tau}+\left[\frac{c \tau}{2}\left(1-\mathrm{e}^{-2 \Delta t / \tau}\right)\right]^{1 / 2} N(t),
\end{equation}
where $\Delta t$ is the time step, $\tau$ the correlation time, $c$ the diffusion constant of the process and $N(t)$ a temporally uncorrelated normally distributed random variable. It is a dimensionless term, which yields an effective Rabi frequency of $\left(1+X \right)\Omega$. Neither of the system error types lead to considerable overheads in a simulation on a quantum computer. Finally, $^{13}C$ nuclear spin decay is not a relevant error source on the time scale of the protocol, since it is of the order of seconds~\cite{13clifetime}, while the hyperpolarization process operates in the order of microseconds.

\section{Co-Design algorithm}
\label{sec:algorithm}

In this section we provide an in-depth description of our Co-Design algorithm, starting with the choice of a simulation technique, followed by a short listing of hardware assumptions related to the allowed qubit operations (gates and resets), as well as the noise and errors present in the physical NMR system and in the quantum computer. Subsequently, the algorithm components are introduced. We end the section with a discussion on layout and gate-level optimization. The high-level structure of the simulation protocol is shown in Fig. \ref{subfig:algorithm_sketch}.

\subsection{Simulation technique}
\label{subsec:simulation_technique}

The best established digital quantum simulation technique is based on decomposing the time-evolution operator into single-qubit and two-qubit gates through the Lie-Trotter-Suzuki formula~\cite{Suzuki1976}, known as Trotterization. To simulate our problem on a quantum computer, we base our strategy on regular Trotterization~\cite{lloyd1996universal} but we also explore the randomized Trotterization method qDRIFT~\cite{Campbell2019} in Appendix~\ref{section:other_sim}. Other, more NISQ-specific, simulation techniques such as the variational quantum simulator~\cite{yuan2019theory}, the quantum assisted simulator~\cite{bharti2021quantum}, numerical quantum circuit synthesis~\cite{younis2021qfast}, and a plethora of other quantum algorithms~\cite{bharti2021noisy} can also be used as simulation methods.

One advantage of Trotterization over some of these NISQ methods is that it closely follows the real time evolution for each time step. This is particularly important for pulsed-driving schemes, where the free evolution in between different pulses always starts with a different initial state. Variational and quantum assisted methods would then require that each interpulse evolution is solved independently, making them impractical for the problem. 

A second advantage of Trotterization is that its complexity and precision are straightforward to analyze. The Trotterization procedure can also be expanded to higher orders, and symmetrized expansions converge more rapidly and reduce the error with respect to the continuum time limit~\cite{Hatano2005}.

\subsection{Hardware assumptions}
\subsubsection{Native gates}
\label{subsubsec:native_gates}
The hardware for the quantum simulation plays a major role in choosing the optimal quantum algorithm and its specific implementation. In our case, we consider a quantum computer based on superconducting qubits with the following native single-qubit gate set:

\begin{eqnarray}
R_{xy}(\phi,\theta)&=&e^{-i(\cos{\phi}X+\sin{\phi}Y)\frac{\theta}{2}}; \,\,\textmd{and}\\\
R_{z}(\theta)&=&e^{-iZ\frac{\theta}{2}},
\end{eqnarray}
where $X$, $Y$, and $Z$ are Pauli operators on the superconducting transmon qubits. The $R_{xy}(\phi,\theta)$ can physically be implemented through a microwave drive~\cite{Krantz2019}. The gate $R_{z}(\theta)$ on the other hand does not need to be implemented directly, but can be performed virtually by tuning the phase of the subsequent gates applied on the qubit~\cite{McKay2017}. This reduces the number of single-qubit gates (SQGs) that need to be implemented.

The native two-qubit gate (TQG) that arises from the superconducting system Hamiltonian shown in Sec.~\ref{sec:co-design_HW} and Appendix~\ref{gate_theory}, is a continuously-parameterized controlled-$Z$ (CZ) interaction~\cite{yan2018tunable}, which can be transformed through local virtual $R_{z}$-rotations into the form of a $ZZ$-interaction:
\begin{align}
\label{ZZ_unitary}
     U_{ZZ}(\phi)  = 
	\left(
	\begin{array}{cccc} 
		e^{-i\phi} & 0 & 0 & 0\\ 
		0 & e^{i\phi} & 0 & 0\\ 
		0 & 0 & e^{i\phi} & 0\\ 
		0 & 0 & 0 & e^{-i\phi}
	\end{array}
	\right).
\end{align}
Even though the $ZZ$-interaction and the controlled-$Z$ interactions appear different, their physical implementation is identical since they are related through virtual $R_{z}$-rotations which come at no additional cost. 

Sec.~\ref{sec:co-design_HW} goes into more depth on the two-qubit-gate implementation on our Co-Design quantum chip. 

\subsubsection{Qubit reset}
In the hyperpolarization process the state of the NV needs to be re-initialized after each cycle. It is therefore necessary to be able to reset the state of the qubit representing the NV center in the quantum computer. A qubit reset operation can be defined by two Kraus operators:
\begin{align}
    K^{\textrm{reset}}_1= \begin{pmatrix}{} 
		1 & 0 \\ 
		0 & 0 \\ 
	\end{pmatrix}, \,
	K^{\textrm{reset}}_2= \begin{pmatrix}{} 
		0 & 1 \\ 
		0 & 0 \\ 
	\end{pmatrix}.
\end{align}

On superconducting hardware this can be realized through connecting a quantum circuit refrigerator (QCR) to each circuit element that needs to be reset~\cite{tan2017quantum,silveri2017theory,hsu2020tunable,sevriuk2019fast}. Different reset schemes are discussed in Sec.~\ref{subsec:HW_reset}.

\subsubsection{Noise and errors}
\label{subsec:noise_and_errors}

In this paper we show that the simulation can tolerate the noise of the quantum processing unit (QPU), and that the simulation does not require large overheads to implement imperfections present in the nanoscale-NMR system, as discussed in Sec.~\ref{sec:hyperpolarization}. We will refer by $\textit{system imperfections}$ to effects in the nanoscale NMR system only, while the QPU is affected by $\textit{noise}$, referring to the effect of the environment on the qubits, and $\textit{errors}$, referring to inaccuracies of gates. 

In our simulation of the algorithm, we use the most common noise models for superconducting transmon qubits~\cite{Krantz2019}, namely an amplitude damping channel modelled by the Kraus operators:
\begin{equation}
\begin{aligned}
&K^{\textrm{amp}}_{1}(t)=|0\rangle\langle 0|+\sqrt{1-p(t)}| 1\rangle\langle 1|=\left(\begin{array}{cc}
1 & 0 \\
0 & \sqrt{1-p(t)}
\end{array}\right), \\
&K^{\textrm{amp}}_{2}(t)=\sqrt{p(t)}|0\rangle\langle 1|=\left(\begin{array}{cc}
0 & \sqrt{p(t)} \\
0 & 0
\end{array}\right),
\end{aligned}
\end{equation}
with $p(t) = 1 - \exp\left(-t/T_1\right)$ and $T_1 = 60\,\mu s$, and
a pure dephasing channel represented by the Kraus operators:

\begin{equation}
K^{\textrm{deph}}_{1}(t)=\left(\begin{array}{cc}
1 & 0 \\
0 & \sqrt{1-p(t)}
\end{array}\right), \, K^{\textrm{deph}}_{2}(t)=\left(\begin{array}{cc}
1 & 0 \\
0 & \sqrt{p(t)}
\end{array}\right),
\end{equation}
with $p(t)=1-\exp (-\Gamma(t))$ and $\Gamma(t)$ given by the expression $\Gamma(t)=\frac{t^{2}}{2} \int_{0}^{\infty} d \omega I(\omega) \operatorname{cotanh}\left(\frac{\beta \omega}{2}\right) \operatorname{sinc}^{2}\left(\frac{\omega t}{2}\right)$ where $\beta$ is the inverse temperature of the environment. We chose the spectral function $I(\omega)$ to be of the type $1/f$~\cite{Krantz2019}, and $T_2 = 60\,\mu s$. Additionally, each gate operation is assumed to be calibrated up to a two-qubit-gate (TQG) error $\varepsilon_{\textrm{TQG}} \in [10^{-4},10^{-2}]$, with the induced effective noise modelled by a depolarizing channel defined for single-qubit gates by the Kraus operators:

\begin{equation}
\begin{aligned}
K^{\textrm{depol}}_{1} &=\sqrt{1-p} \ I, \\
K^{\textrm{depol}}_{2} &=\sqrt{p / 3} \ X, \\
K^{\textrm{depol}}_{3} &=\sqrt{p / 3} \ Y, \\
K^{\textrm{depol}}_{4} &=\sqrt{p / 3} \ Z,
\end{aligned}
\end{equation}
and for two-qubit gates by an analogous expression with the tensor products of two Pauli matrices and the coefficients $\sqrt{1-p}$ for the identity and $\sqrt{p/15}$ for the other operators.

Single-qubit-gate (SQG) errors $\varepsilon_{\textrm{SQG}}$ are assumed to be one order of magnitude lower than TQG errors.

\subsection{Algorithm components}

\begin{figure*}[h!btp]
    \subfloat[\label{subfig:algorithm_sketch}]{%
        \includegraphics[width=0.99\linewidth]{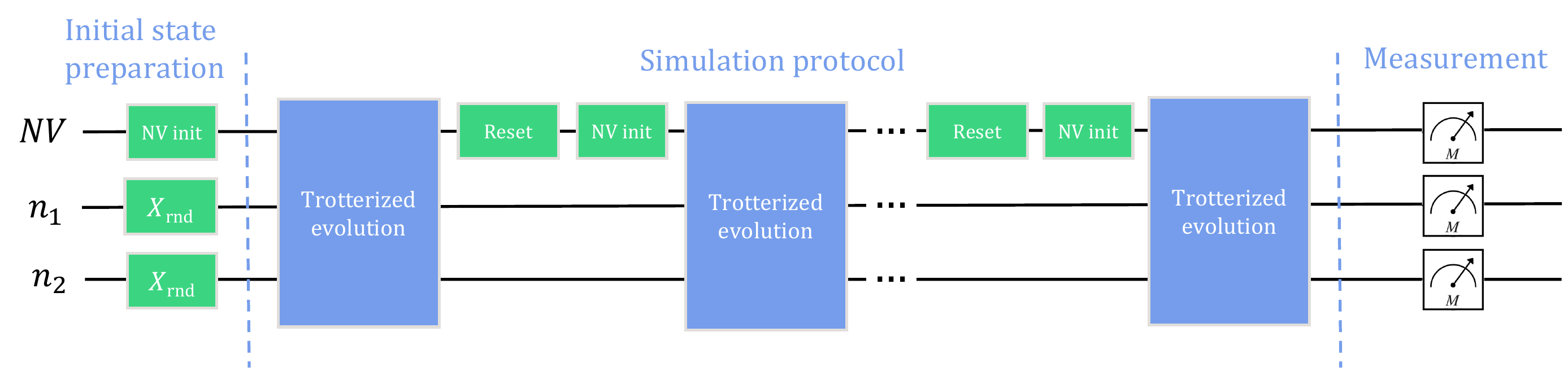}
    }\hfill
    \centering
    \subfloat[\label{subfig:one_trotter_step}]{%
        \includegraphics[width=0.99\linewidth]{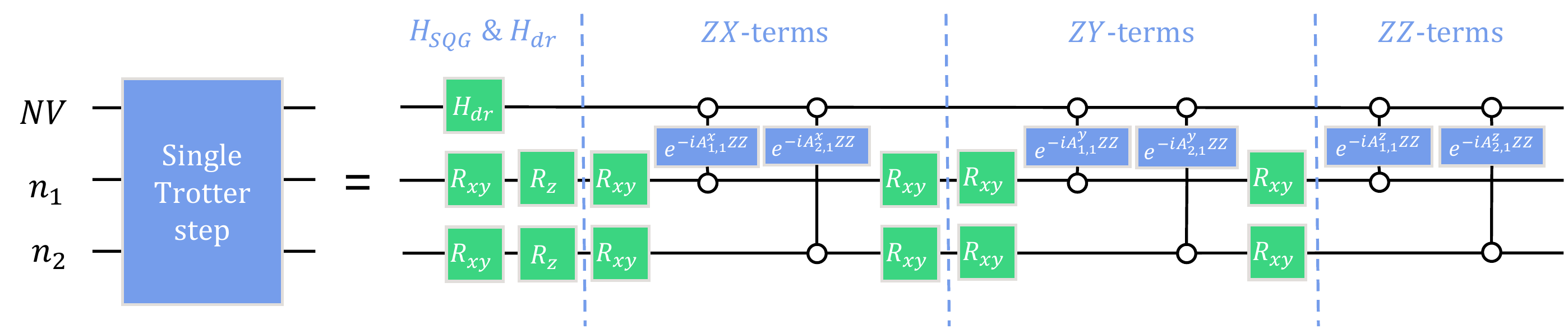} 
    }\hfill
    \caption{(a) Sketch of the overall operation of the simulation algorithm for one NV center and two nuclei, with continuous driving; (b) corresponding gate sequence of one Trotter step on a star-topology chip for non-interacting nuclei. $H_{\textrm{SQG}}$ refers to the single-qubit-gate component of the Hamiltonian, $A_{j,1}^{x,y,z}$ parameters are the various coupling strengths of the simulated system, and the $X_{\textrm{rnd}}$ gates refer to $X$-gates applied with a 50\% probability to prepare an effective fully-mixed state. The initial state preparation can also be performed using the alternative random-phase approximation-inspired method. NV init is an initial-state preparation using single-qubit gates to the state required by the driving scheme. Details of the circuit components can be found in Appendix \ref{section:appendix_time_evol} along with a figure representing the pulsed driving case.
    }\label{fig:circuit_comparison_graphics}
\end{figure*}

Our simulation of the nanoscale NMR problem follows the general structure shown in Fig.~\ref{subfig:algorithm_sketch}. It starts by initializing the states of all qubits, according to whether they represent a nucleus or a NV center, then evolving them using Trotter steps, followed by reset and re-initialization of the qubits representing NV centers. The cycle of time evolution and re-initialization is then repeated as many times as the protocol calls for. Finally the qubits are measured, and the polarization of the NV centers and nuclei are extracted as the expectation values of the qubit representing each element. Fig.~\ref{subfig:algorithm_sketch} shows the circuits for the case of continuous driving, while the details of pulsed driving schemes are shown in Fig.~\ref{subfig:algorithm_sketch_pulsed} in Appendix~\ref{HH_sequence}. In the following, we go through these steps in more detail for the case of a single NV center. 

\subsubsection{Initial state preparation}

To enable the polarization transfer, it is necessary to prepare the NV center in a specific initial state that depends on the driving scheme. For the continuous-driving scheme it is the $|+\rangle$ or $|-\rangle$ state, and for the pulsed-driving scheme it is one of the two computational basis states, $|0\rangle$ or $|1\rangle$.

For a diamond at room temperature, the initial state of the nuclear spins is well described by a fully-mixed state $\rho_{\textrm{mixed}}=\frac{\mathbb{1}^{\otimes N}}{2^{N}}$, where $\mathbb{1}^{\otimes N}$ is the $2^N \times 2^N$ identity matrix. The state can be approximated by running the algorithm several times, each time with a different initial state obtained by applying $X$ gates randomly on the qubits representing nuclei. A faster alternative to this sampling is the random-phase-approximation-inspired method, described in~\cite{celio1986new}, and introduced into quantum computing in~\cite{McArdle2021}. In this method, the qubits are all prepared in an equal superposition by applying Hadamard gates, and then the phases are randomized through the application of random phase gates. The method effectively reduces the prefactor in the scaling of the sampling error~\cite{McArdle2021}. 

\subsubsection{Time evolution}
\label{subsec:time_evol}
We choose to implement the time evolution generated by the Hamiltonian in Eq.~\eqref{eq:hamiltonian1} through Trotterization. For that, the Hamiltonian is rewritten in terms of qubit Pauli operators and arranged into non-commuting terms for an optimal Trotter splitting. The resulting circuit, which performs one Trotter step of the evolution in the continuous driving case, is depicted in Fig.~\ref{subfig:one_trotter_step}. It consists of a set of initial single-qubit gates, including the ones corresponding to the driving and the detuning of the NV center, followed by three two-qubit gates per nucleus. There are three types of interaction terms, of the form $XZ$, $YZ$ and $ZZ$, when no internuclear interactions are considered. With interactions there are a total of five interaction terms. Our native gate set only includes one type of two-qubit interaction as explained in section~\ref{subsubsec:native_gates}. 

Therefore, some SQGs need to be applied in order to convert the interaction terms into the right form, as discussed in Appendix~\ref{section:appendix_time_evol}.

Under specific circumstances, some TQGs can be removed by rotating the Hamiltonian into a more suitable basis as shown in Appendix~\ref{section:rotational_opt}.

\subsubsection{Cycles and reset}

The dynamics of the system is known to produce an exchange of polarization between the NV center and the nuclei. This exchange is oscillatory, and therefore choosing a proper stopping time is important in order to achieve an effective polarization transfer from the NV center to the nuclei. In practice, a sub-optimal transfer time can suffice, and the protocol is then repeated several times by resetting the NV center to its initial state and letting the system evolve under the drive again. Due to the re-initializations the full evolution of the system is non-unitary and a net gain of polarization of the system is enabled. 

This structure is represented in the quantum circuit in Fig.~\ref{subfig:algorithm_sketch} by the repeated Trotter evolution, followed by reset operations on the qubit representing the NV center, and a single-qubit gate to prepare the initial state of the driving protocol.  

\subsection{\label{subsection:layout}Layout optimization}

When implementing a quantum algorithm on a superconducting QPU, the planar qubit connectivity forces us to solve the qubit-routing problem by introducing additional SWAP gates to connect distant qubits. In this subsection, we study the advantages of an optimized chip topology, a star topology, over a square-grid array of qubits in terms of reducing the number of SWAP gates that must be inserted to run the algorithm in  Fig.~\ref{fig:circuit_comparison_graphics} on the device. 

\begin{figure}[h!btp]
\begin{center}
    \subfloat[\label{fig:swap_square}]{%
        \includegraphics[width=0.95\linewidth]{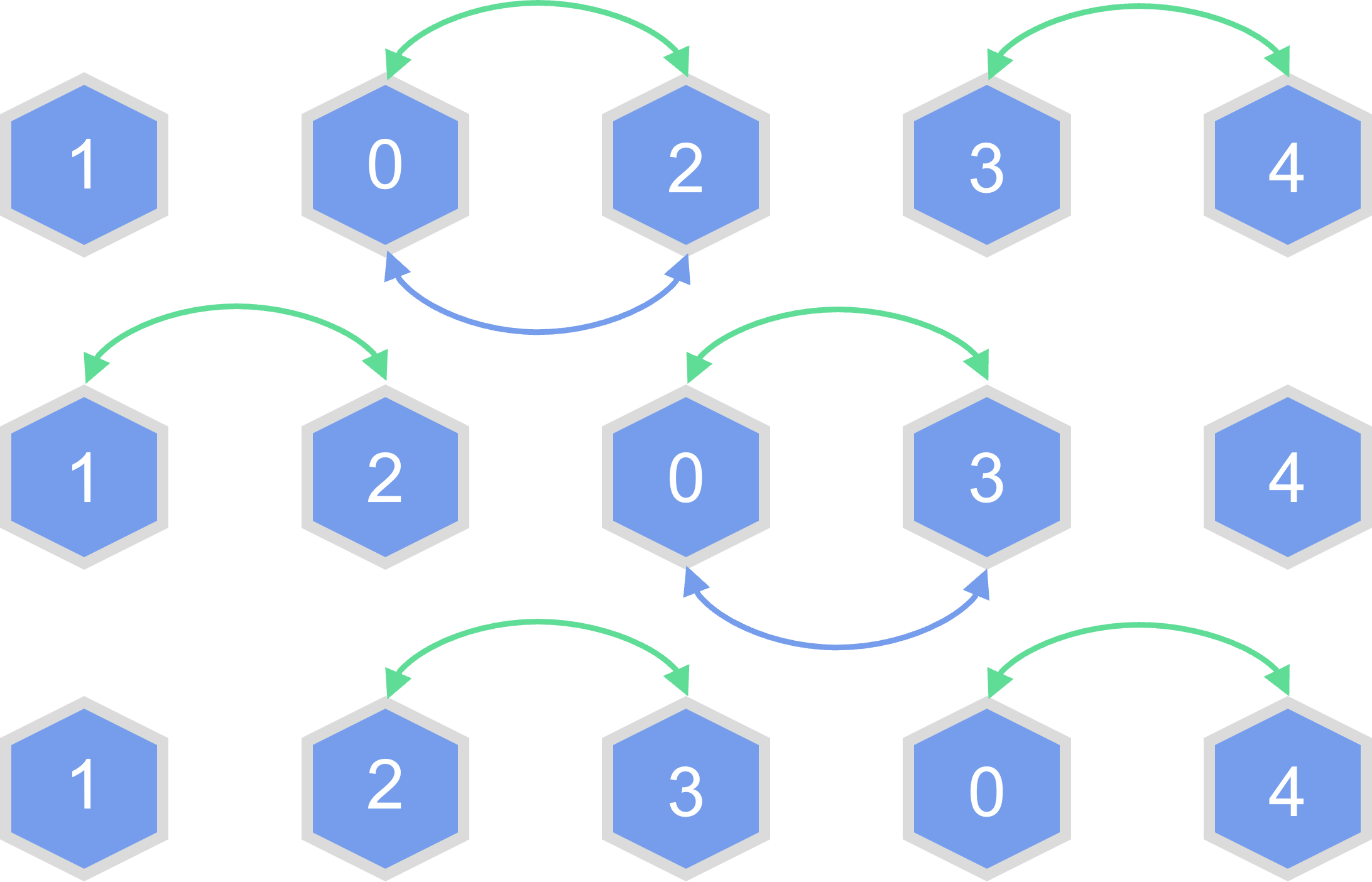} 
    }\hfill
    \subfloat[\label{fig:star_chip_topology}]{%
        \includegraphics[width=0.66\linewidth]{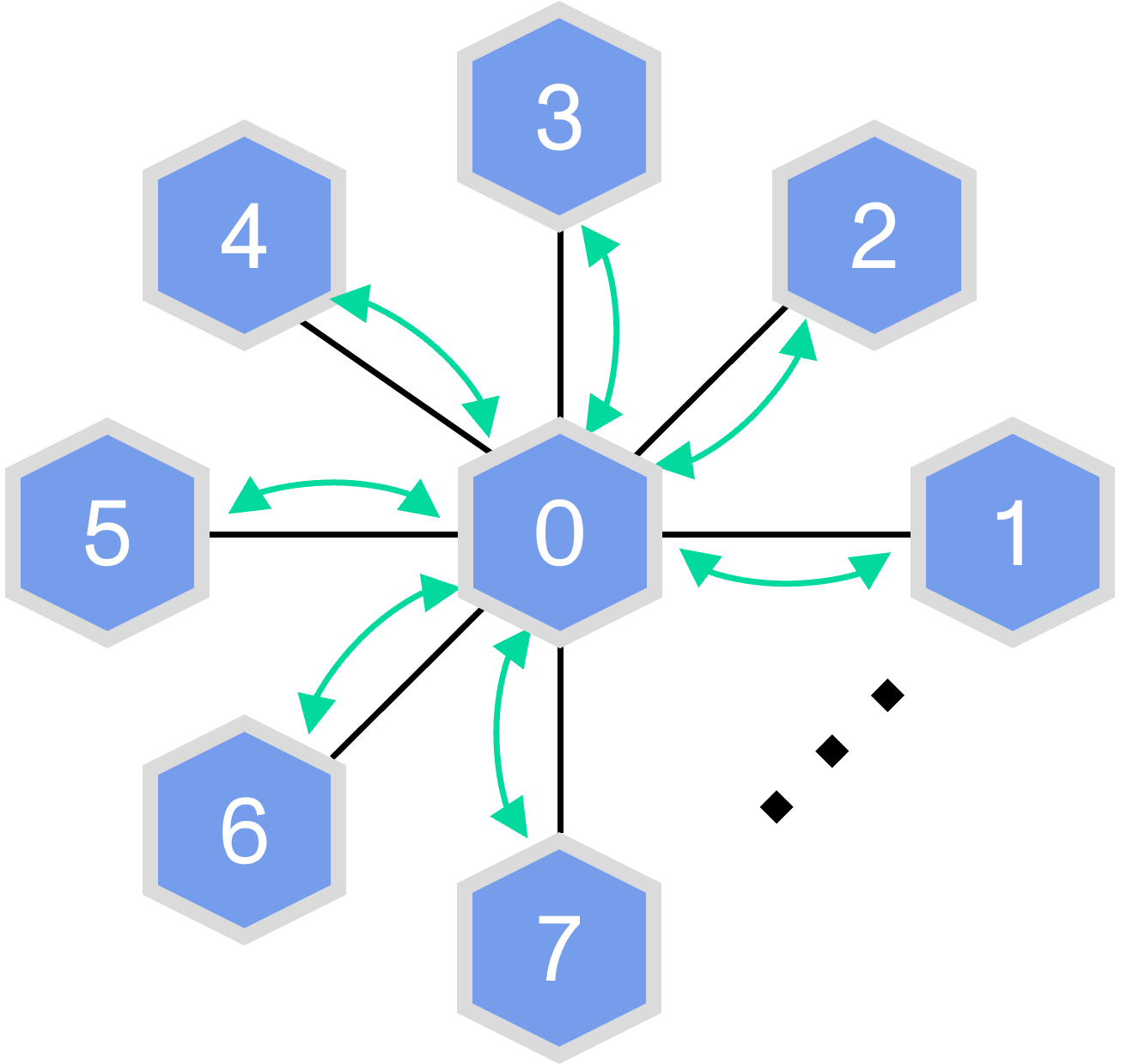}
    }\hfill

    \caption{(a) Three steps of the SWAP patterns in a five-qubit linear chain displayed from top to bottom. Green (blue) arrows represent the SWAP pattern for the case with (without) internuclear interactions. The green pattern is known as the `odd-even' SWAP pattern. The numbers are expressed according to the blue pattern, where label 0 represents the position of the NV center. (b) Star chip topology with the SWAP pattern for the interaction with internuclear interactions. }\label{fig:swap_patterns}
    \end{center}
\end{figure}
Different topologies will imply different counts of SWAPs added on top of the gates arising from the algorithm itself, as shown in Fig.~\ref{fig:swap_patterns}. On a NISQ device, this implies different computational precision for the same gate error magnitudes.
We choose the SWAP count as our metric to compare different topologies, as commonly gates have fidelities limited by calibration. The errors could be due to crosstalk, leakage, or filtering causing disturbances to the control signals. Under this scenario we want to minimize the gate count. On the other hand, for a highly tuned up device whose gates are limited by qubit coherence times, it would be optimal to minimize the circuit depth instead of the TQG count.

Assuming the gate errors are independent, the total error will be bounded by:
\begin{equation}
\varepsilon_{\textrm{gates}}=1- {(1-\varepsilon_{\textrm{TQG}})} ^{N_{\textrm{TQG}}} (1-\varepsilon_{\textrm{SQG}})^{ N_{\textrm{SQG}}}, 
\end{equation}
where $N_{\textrm{TQG}}$ is the number of two-qubit gates, $N_{\textrm{SQG}}$ the number of single-qubit gates, and $\varepsilon_{\textrm{SQG}}$ is the SQG error. 
Consequently, reducing the gate count, especially $N_{\textrm{TQG}}$, has an exponential effect on the precision of the computations, underlining the effect of minimizing the SWAP gate overhead. As SWAP gates are not native to the hardware, but must be compiled out of three CZ gates, their effective error rate is also much higher than those of native gates.

\subsubsection{Square grid}

A common choice in superconducting quantum chips is the square grid of qubits. It has high connectivity and is suitable for performing the surface code error correction when scaled to large enough qubit counts with fast measurement and feedback~\cite{Fowler2012}. 
The qubit routing problem on a square grid can be tackled using various numerical approaches~\cite{Hirata2009,Li2019,Saeedi2010,Zulehner2018}. However, these methods are inefficient. In our case, a tailored SWAP routing method, shown in Fig.~\ref{fig:swap_square}, has been chosen and developed in Appendix~\ref{sec:routing} that can be shown to be well suited from two perspectives. First, a comparison against the cited numerical approaches (shown in Appendix~\ref{sec:routing}) reveals that our routing method is better in terms of number of gates. Second, it is completely deterministic and does not rely on expensive numerical optimization methods. It can also be shown not to be far from optimal: on a square grid each qubit has at most 4 nearest neighbors, implying that any SWAP operation provides at most 3 new neighbors. For an all-to-all (ATA) interacting Hamiltonian there are $\frac{n^2}{2}$ interactions, to leading order, for a simulation performed on $n$ qubits (corresponding to $N$ nuclei and one NV center). This implies a lower bound of at least $\frac{n^2}{6}$ SWAPs for any SWAP pattern on the square grid topology. Our SWAP pattern with $\frac{n^2}{2}$ SWAPs, discussed in Appendix~\ref{sec:routing}, is thus not far from optimal.

\subsubsection{Star architecture}

\begin{figure}[h!btp]
\centering
    \includegraphics[width=0.99\linewidth]{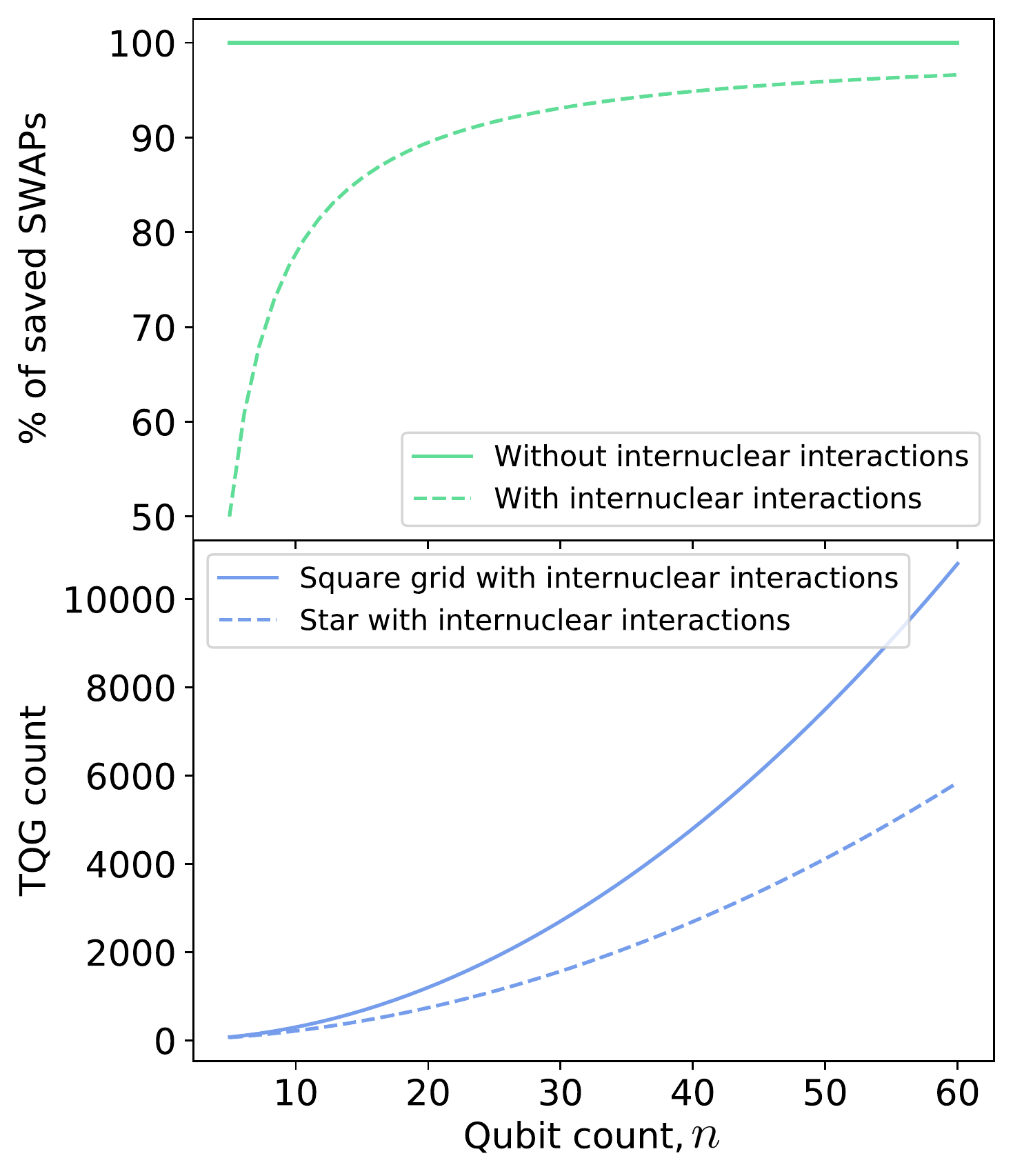} 
    \hfill
    \caption{The (top) panel shows the percentage of SWAP gates saved by using a star topology instead of a square grid for $n$ qubits for the cases with and without internuclear interactions. The (bottom) panel shows  the total TQG count against the qubit count in the interacting case for the square grid and the star architecture.}\label{fig:TQG_count_and_saved_swaps.pdf}
\end{figure}

A star topology allows to implement the simulation of the simplified case without internuclear interactions directly, without any SWAP gates. With internuclear interactions considered, we still find a reduction in SWAP gates as compared to the square grid topology, as shown in Fig.~\ref{fig:star_chip_topology}. This reduction comes from the SWAP routing we implement, that consists of making the qubit $0$ in Fig.~\ref{fig:star_chip_topology} interact with all the external qubits and then swap its state with that of qubit $1$ and repeat this process until all interactions have been performed. This allows us to use only $n-1$ SWAP gates. The percentage of SWAP gates that can be saved can be observed in Fig.~\ref{fig:TQG_count_and_saved_swaps.pdf}. 

However, this improvement in the number of gates comes with a price to pay in the depth of the algorithm. We can only do one TQG at a time in the star chip and we have $\frac{3}{2}n(n-1)$ TQGs from simulating the physical interactions and $3(n-2)$ TQGs from the SWAPs. This yields a depth for the TQGs of $\frac{3}{2}n^{2}+\frac{3}{2}n-6$ in a star chip, while for a square grid it is $6n$. Such depth increase comes from the reduction in parallelization, since all gates now act via the central qubit. On the other hand, less parallelization reduces the types of possible crosstalk errors. Adding connections between external qubits reduces the depth of the circuit, since the main cause of circuit depth is the fact that the interaction of two external qubits needs to be done exclusively by the central qubit. Further studies are required to see if the addition of more external layers to this topology (such as in a spiderweb) can lead to better compromises between depth and gate count, especially for simulating systems with clusters of strongly interacting nuclei.

\subsection{Gate-level optimization}
The two-qubit interactions that appear in the algorithm are the $XZ$, $YZ$ and $ZZ$ interactions, as shown in section~\ref{subsec:time_evol} and Fig.~\ref{subfig:one_trotter_step}. When compiling the algorithm into the native gates of the device, all these interactions must be implemented in terms of some available gate set. We study in Table~\ref{tab:gates-overhead} the overhead introduced by decomposing these interactions into different examples of native TQGs of superconducting devices; namely, the parametrizable and fixed-phase $U_{ZZ}$ gate, the fixed-phase controlled-$Z$ gate $\mathrm{CZ}$, and the $\mathrm{CNOT}$ gate. The $\mathrm{CNOT}$ gate is usually performed by making use of the cross-resonance gate \cite{Krantz2019, magesan2020crossresonance}, which introduces an $U_{XZ}$ interaction, making it equivalent to the $U_{ZZ}$ for the purpose of this algorithm. We assume that the SQGs that can be implemented are the $R_{xy}$ and the $R_{z}$ gates. These numbers can be further reduced if the first and last SQGs introduced by this compilation are combined with the adjacent SQGs in the algorithm.

\begin{table}[]
\begin{tabular}{|l|c|c|c|c|}
\hline
                     & $U_{ZZ}(\phi)$ & $U_{ZZ}(-\pi/4)$ & $\mathrm{CZ}(\pi)$ & CNOT \\ \hline
\textbf{TQGs}        & $1$         & $2$     & $2$        & $2$    \\ \hline
\textbf{SQGs}        & $0$         & $5$          & $3$        & $1$    \\ \hline
\end{tabular}%
\caption{Overheads introduced by the decomposition of $U_{ZZ}(\phi)$ gates into different examples of native TQGs in superconducting devices. The single-qubit-gate (SQG) count includes only $R_{xy}$ rotations, as the $R_{z}$ rotations can be implemented virtually.}
\label{tab:gates-overhead}
\end{table}

The conclusion is that fixed-angle gates will double the number of TQGs that need to be physically performed. 
In Ref.~\cite{lacroix2020continuousgates}, the improvements coming from the reduction of the gate count are compared to the new errors introduced by the interpolation of the calibrated phases. For two instances of a Quantum Approximate Optimization Algorithm (QAOA) \cite{farhi}, it is shown that the performance is better when using parametrized TQGs. 

\begin{figure}[h!btp]
\begin{center}
    \subfloat[\label{fig:gate_decomp1}]{%
        \includegraphics[width=1\linewidth]{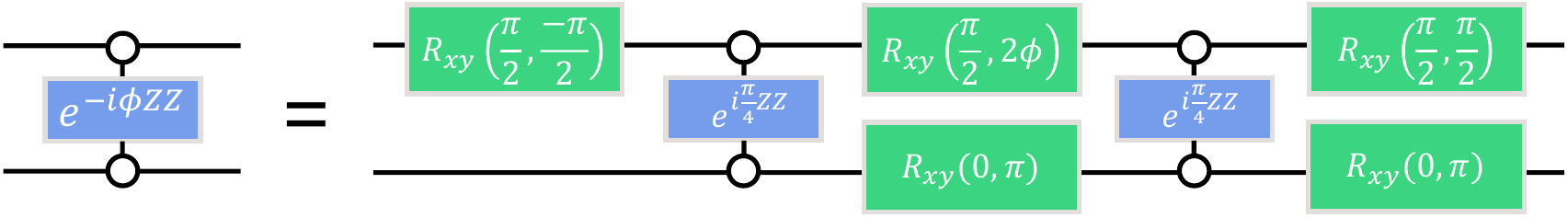} 
    }\hfill

    \subfloat[\label{fig:gate_decomp3}]{%
        \includegraphics[width=0.75\linewidth]{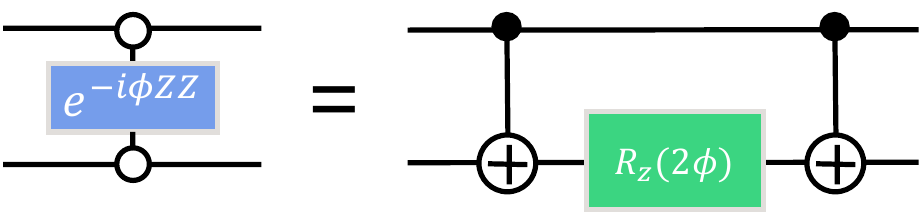}
    }\hfill

    \caption{(a) Gate decomposition of $e^{-i\phi ZZ}$ in terms of the fixed-phase $U_{ZZ}(\frac{\pi}{4})$ gate, (b) the $CNOT$. }\label{fig:gate_decomp}
    \end{center}
\end{figure}

The gate sequences for some of the gate decompositions are shown in Fig.~\ref{fig:gate_decomp}.

\section{Co-Design hardware}
\label{sec:co-design_HW}

\begin{figure}
\subfloat[\label{subfig:6donis}]{%
        \centering 
        \includegraphics[width = 0.8\linewidth]{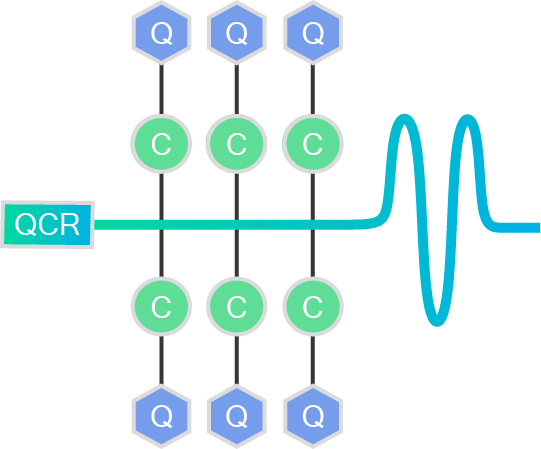}
    }

\subfloat[\label{subfig:QCR_circuit}]{%
        \centering 
        \includegraphics[width = 0.9\linewidth]{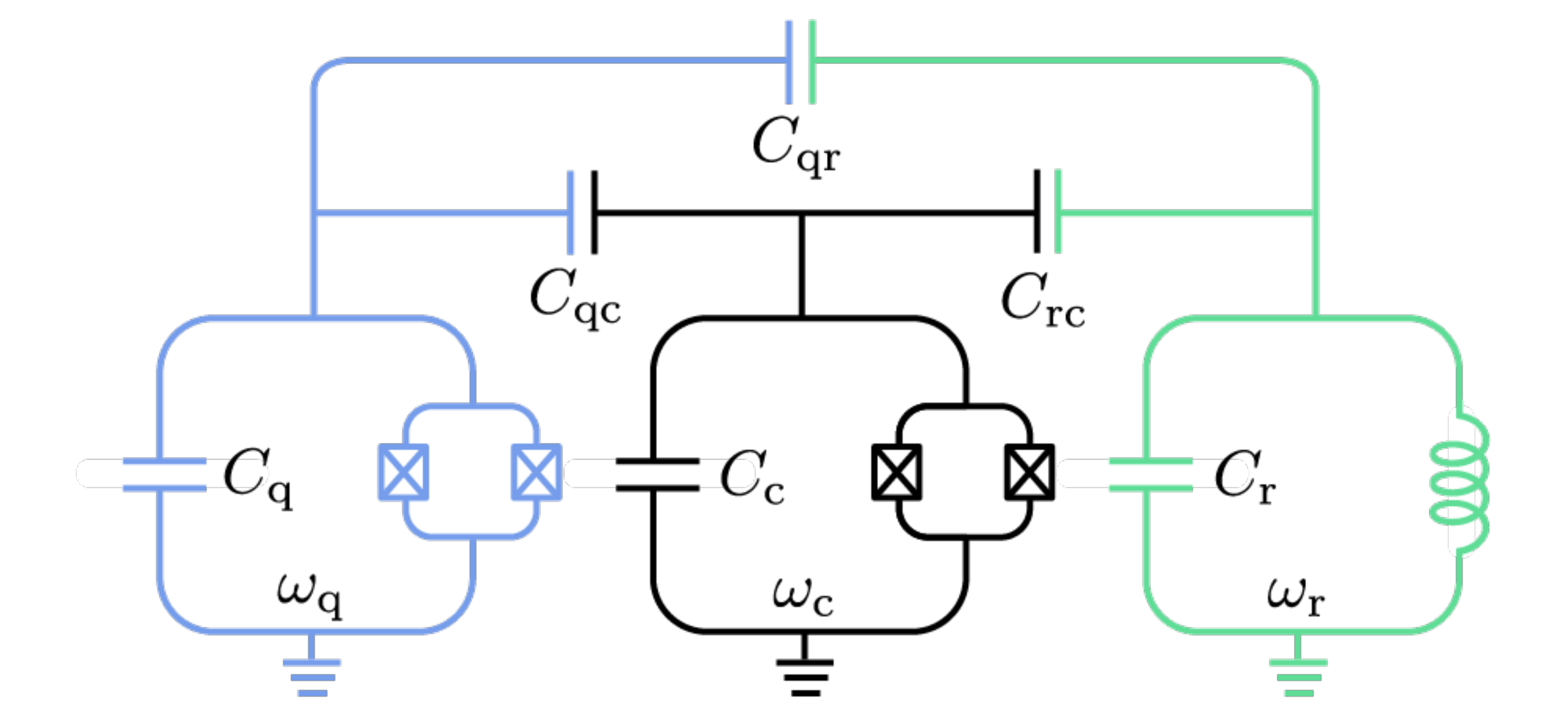}
    }\hfill
\caption{(a) Central $\lambda/4$ resonator with 6 qubits coupled via tunable couplers. The resonator is also coupled to a quantum circuit refrigerator enabling fast reset. The device acts effectively as a 6 qubit star-architecture chip.\\
         (b) Electrical diagram of transmon qubit (left) coupled to a resonator mode (right) via a tunable coupler (center). The qubit has frequency $\omega_\textrm{q}$, coupler $\omega_\textrm{c}$, and resonator $\omega_r$. The qubit and resonator have a direct capacitance $C_\textrm{qr}$ and capacitances $C_\textrm{qc}$ and $C_\textrm{rc}$ respectively to the coupler.} 
         \label{fig:ndonis}
\end{figure}

A star-architecture chip has fundamental scaling issues using a transmon as the central qubit as the number of neighbors grows. Every neighbor added to the center qubit would decrease its charging energy $E_{c}$. To keep the qubit frequency constant and anharmonicity in the transmon regime, the ratio of the qubit's Josephson energy to its charging energy, $E_j/E_c$, must remain unaffected. Therefore we cannot afford to change its charging energy. This leads to a trade-off between the number of coupled qubits and their coupling strength to the central element.

The spirit of Co-Design calls for replacing the central transmon with another object that enables this scaling in size. A resonator has no Josephson energy $E_{j}$, so the $E_j/E_c$ ratio is not altered by adding more capacitive couplings to the resonator. Only small corrections to its frequency are introduced by adding coupled qubits. As a distributed element, a co-planar waveguide resonator also has physically more space for couplings than a central transmon qubit. By elongating the resonator and choosing the mode with the target frequency, the number of qubits coupled to it can further be increased. These properties make a resonator a favourable component in the center of the chip.

In the device in Fig.~\ref{subfig:6donis} the qubits are capacitively coupled to the resonator via tunable couplers~\cite{mariantoni2008two,yan2018tunable,foxen2020demonstrating} in the proximity of a voltage maximum of a standing wave in the resonator. As the resonator is elongated, we must use higher harmonic excitations of the resonator to keep the frequency around the operational frequency of the qubits. Tunable couplers avoid the frequency crowding issues related to direct coupling~\cite{song201710,song2019generation}, and the linear resonator has higher connectivity in the center than ring resonator structures with quasi-all-to-all connectivities~\cite{hazra2021ring}.

A linear resonator cannot in general be used as a qubit, since a microwave drive on it will not only populate the $\{\ket{0},\ket{1} \}$ subspace, but also higher excited states. However, the effective interactions mediated via the tuneable coupler in Fig.~\ref{subfig:6donis} are of the type $a^{\dagger}a Z$ and $(a+a^{\dagger})X+(a-a^{\dagger})Y$ where $a$ and $a^{\dagger}$~\cite{Krantz2019} are the resonator creation and annihilation operators. These types of interactions conserve excitation number, so when at most one excitation is in the qubit-resonator system, the resonator cannot be populated beyond its first excited state through interaction with a qubit mediated a tuneable coupler.

CZ and iSWAP gates between the resonator and a qubit can be performed using the two interactions, and the theory is developed more fully in Sec.~\ref{subsec:gate_theory}. Then, a resonator together with an external qubit can be used as an effective central qubit in the following way: 

\begin{enumerate}
\item Prepare all qubits and the resonator in their ground states
\item Select one qubit to form the effective central qubit together with the resonator

\item Prepare an arbitrary state in the selected qubit
\item Perform an iSWAP operation from the selected qubit to the resonator initially in the ground state
\item Perform CZ gates between the resonator and any other qubits
\item Perform an iSWAP operation back from the resonator to the selected qubit for measurement
\end{enumerate}
The theoretically most straightforward protocol would be to perform a SWAP gate from the qubit to the resonator. The iSWAP, on the other hand, is a native gate that can directly be implemented on the hardware in Fig.~\ref{subfig:QCR_circuit}. The iSWAP gate between the resonator and the qubit is represented by the unitary operator:

\begin{align}
    U_{\mathrm{iSWAP}} = 
	\left(
	\begin{array}{cccc} 
		1 & 0 & 0 & 0\\ 
		0 & 0 & -i & 0\\ 
		0 & -i & 0 & 0\\ 
		0 & 0 & 0 & 1 
	\end{array}
	\right).
	\label{iSWAP_unitary}
\end{align}

Since the CZ gates performing the computation following the iSWAP are diagonal in the computational basis, the phase introduced by the iSWAP is uninvolved in the gate. This enables substituting the SWAP gate by an iSWAP gate in the protocol to further minimize the gate count. 

\subsection{Gate theory and simulations}
\label{subsec:gate_theory}

Here we demonstrate that in our star architecture CZ and iSWAP-type gates between any of the qubits and the $\{\ket{0},\ket{1} \}$ subspace of a chosen resonator mode can be implemented. The operational principles of these gates are very similar to those between two qubits coupled with a tunable coupler~\cite{mariantoni2008two,yan2018tunable,foxen2020demonstrating, chu2021coupler}. The main limitation of our architecture (where one transmon is replaced by a resonator) is that iSWAP operations can only be performed in the zero- and single-excitation subspace of the two-qubit computational basis. 

\begin{table}[H]
\centering
\begin{tabular}{ |c|c|c| } 
 \hline
 Parameter & Symbol & Value \\ 
 \hline
 Resonator frequency & $\omega_r$ & 2$\pi \times$4.3 GHz \\ 
 Qubit anharmonicity & $\alpha_q$ & - 2$\pi \times$0.187 GHz \\ 
 Coupler anharmonicity & $\alpha_c$ & - 2$\pi \times$0.110 GHz \\ 
 Resonator-coupler coupling & $g_{rc}$ & 2$\pi \times$98.5 MHz \\
 Qubit-coupler coupling & $g_{qc}$ & 2$\pi \times$101.8 MHz \\
 Resonator-qubit coupling & $g_{rq}$ & 2$\pi \times$8.9 MHz \\
 Resonator relaxation & $T_1^r$ & 60 $\mu s$ \\ 
 Qubit relaxation & $T_1^q$ & 60 $\mu s$ \\ 
 Coupler relaxation & $T_1^c$ & 30 $\mu s$ \\ 
 Resonator dephasing & $T_2^r$ & 60 $\mu s$ \\ 
 Qubit dephasing & $T_2^q$ & 60 $\mu s$ \\ 
 Coupler dephasing & $T_2^c$ & 30 $\mu s$ \\ 
 \hline
\end{tabular}
\caption{Parameters of star-architecture chip.}
\label{tab:ndonis_parameters}
\end{table}

\subsubsection{Conditional-Z gate}

The CZ operation between the resonator and the qubit is described by the unitary operator: 

\begin{align}
\label{CPHASE_unitary}
    \textrm{CZ}(\phi) =
	\left(
	\begin{array}{cccc} 
		1 & 0 & 0 & 0\\ 
		0 & 1 & 0 & 0\\ 
		0 & 0 & 1 & 0\\ 
		0 & 0 & 0 & e^{-i\phi} 
	\end{array}
	\right).
\end{align}

This gate is equivalent to the $U_{ZZ}(\phi)$ gate in Eq.~\ref{ZZ_unitary} up to two $R_z$-rotations. To operate a CZ gate, we initialize the resonator-coupler-qubit set up shown in Fig.~\ref{subfig:QCR_circuit} at the idling configuration with zero effecting coupling between the qubit and resonator. Note that the coupler is also a transmon that shows a higher sensitivity to the magnetic flux than regular qubits. 
We next apply a flux pulse that lowers the coupler frequency, turning on the effective coupling between the resonator and the qubit. Depending on the flux pulse shape, the state collects conditional phase $\phi$ and possibly experiences population oscillations between computational and non-computational states, as a function of the time spent at the gate-operation frequency. We optimize the pulse amplitude and duration such that after the flux pulse the CZ gate fidelity is maximized.
Details of the gate theory can be found in Appendix~\ref{gate_theory} and the considered device parameters in Table~\ref{tab:ndonis_parameters}.

\begin{figure}
\subfloat[\label{fig:CPHASE}]{%
        \centering 
        \includegraphics[width = 0.9\linewidth]{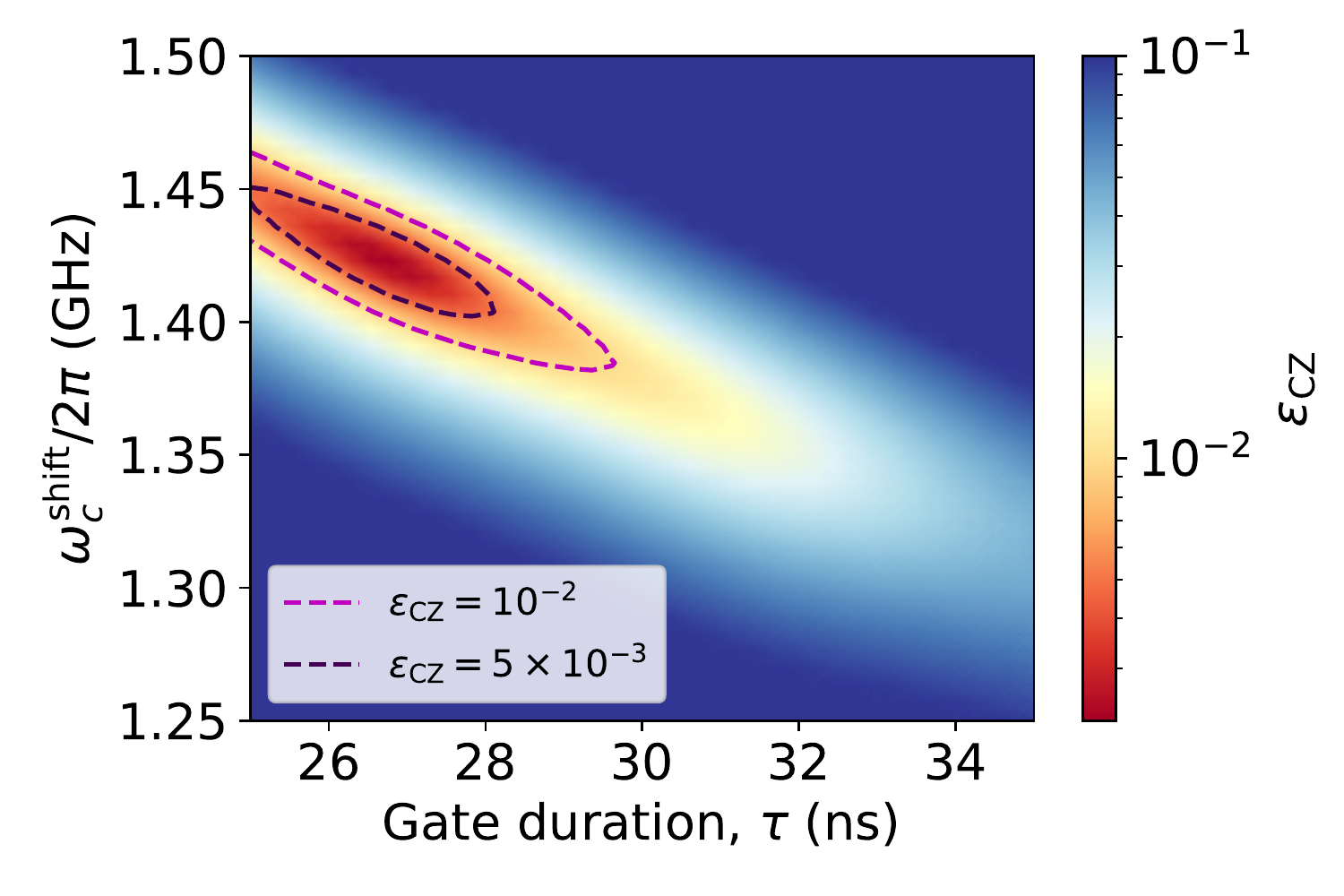}
    }

\subfloat[\label{fig:iswap}]{%
        \centering 
        \includegraphics[width = 0.9\linewidth]{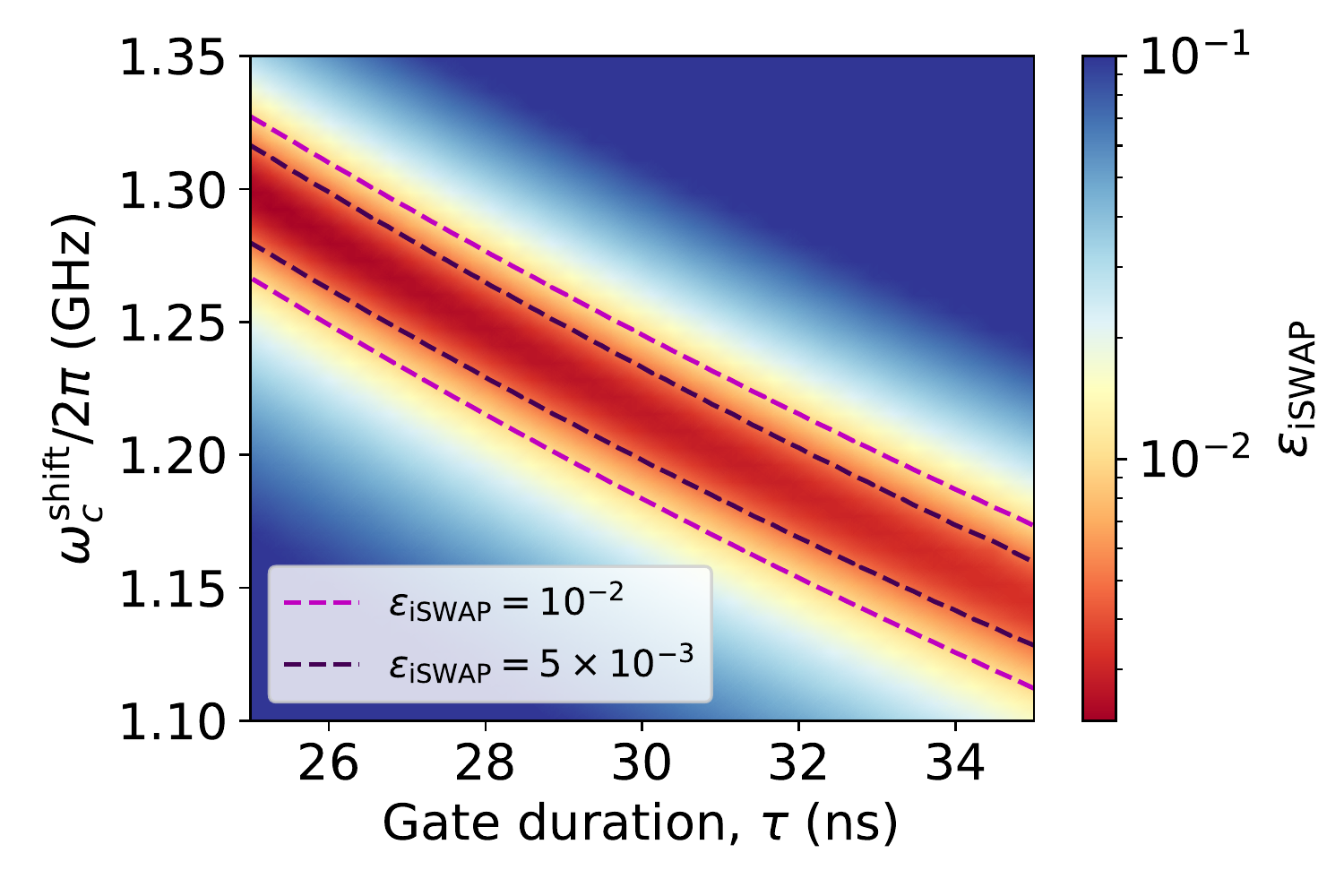}
    }\hfill
\caption{(a) CZ gate error landscape averaged over random initial states.  Contours with a low error are highlighted  with a dashed line. (b) iSWAP gate error landscape obtained by averaging over a number of random initial states in the zero- and one-excitation manifolds.  Both plots are produced using system parameters shown in Table~\ref{tab:ndonis_parameters}.}
         \label{fig:CPHASE_iSWAP}
\end{figure}

In Fig.~\ref{fig:CPHASE}, we operate our CZ gate by tuning the coupler frequency using a flattop Gaussian shaped flux pulse. The width of our Gaussian filter was fixed at 3 ns. Applying such a flux pulse to coupler results in a coupler frequency shift by $\omega_c^{\mathrm{shift}}$ from the idling configuration. Then by appropriately tuning $\omega_c^{\mathrm{shift}}$ and the gate time $\tau$, one locates the optimal pulse configuration that minimizes the CZ($\pi$) gate error $\varepsilon_{\textrm{CZ}} =1 - \big( \mathrm{tr}\sqrt{\sqrt{\rho}\sigma \sqrt{\rho}}\big)^2$, where $\sigma$ is the target density matrix obtained after propagating some initial state $|\Psi \rangle \langle\Psi|$ with the ideal unitary of Eq.~\ref{CPHASE_unitary} and $\rho$ the final density matrix obtained after propagating $|\Psi \rangle \langle\Psi|$ with the Lindbladian corresponding to our system defined in Eq.~\eqref{H2QG}. For our device parameters, the maximal decoherence limited CZ gate error averaged over a number of random initial states is $1.6 \times 10^{-3}$. Note that the system parameters in Table~\ref{tab:ndonis_parameters} were chosen such that they allow for the possibility to find a good idling configuration, where the residual CZ interaction vanishes before the gate operation. In our simulations, we have included environmental noise, such as amplitude damping and pure dephasing and treated them using a Lindblad master equation solver in QuTiP \cite{qutip, qutip2}.

\subsubsection{iSWAP gate}
Just as the CZ gate, the iSWAP gate can be natively realized in superconducting quantum computing architecture~\cite{Krantz2019}. With our device, we can perform high-fidelity iSWAP gates between zero- and single-excitation computational states. The two-photon state $|\rm 1\rangle_r \otimes |\rm 1\rangle$, where $|\rm 1\rangle_r$ denotes the first excited state of the resonator, must be excluded because it resonantly interacts with the state $|\rm 2\rangle_r \otimes |\rm 0\rangle$ inducing a population exchange between the states. Hence the resulting operation in this subspace does not match the action of the targeted iSWAP operation. 

The capacitive coupling between the elements of the electrical circuit shown in Fig.~\ref{subfig:QCR_circuit} gives rise to an effective $XY$-interaction between the qubit and resonator under the rotating wave approximation. Such an interaction conserves excitation number. With only the qubit or resonator (or neither) initially populated, we stay within the single excitation subspace of the joint system, thereby minimizing leakage of quantum population into the higher excited states of the resonator. 
The $XY$-interaction can be turned on by first tuning the qubit in resonance with the resonator, and then applying a flux-pulse to the coupler to turn on the coupling, similar to the CZ gate operation.

Fig.~\ref{fig:iswap} shows iSWAP gate error landscape for the same device parameters (given in Table~\ref{tab:ndonis_parameters}). The optimal average iSWAP gate error $\varepsilon_{\mathrm{iSWAP}}$ obtained for our device is $1.7 \times 10^{-3}$. This result is obtained by averaging over a number of random initial states within the zero- and one-excitation manifolds. 

The results of our two-qubit-gate simulations demonstrate that our star architecture supports operating gates with similar fidelities as regular transmon qubits coupled together. The increased local connectivity of the device reduces the need for SWAP gates to simulate the nanoscale NMR problem (and others with a similar structure) and consequently in the end improves simulation fidelities.

\subsection{Reset}
\label{subsec:HW_reset}
The hyperpolarization protocol described in Sec.~\ref{sec:hyperpolarization} needs regular re-initializations of the state of the NV center. The Co-Design hardware for simulating the protocol must therefore support this operation within qubit lifetimes. This is a hardware challenge, but one with solutions in sight. In particular, the quantum circuit refrigerator (QCR) has been used to perform the reset in tens of nanoseconds~\cite{tan2017quantum,silveri2017theory,hsu2020tunable,sevriuk2019fast}, which is a similar timescale to gate operations. The advantage of using a QCR for the reset is the possibility to reset the central resonator directly, without the need transfer the resonator population back to the central qubit using an iSWAP gate. Alternatively, a fast reset is possible through applying a flux drive to a qubit to SWAP its state with its measurement line~\cite{zhou2021rapid}. This scheme has the advantage of not requiring any additional hardware not already present on the chip, but comes with a small cost in the circuit depth, as the state of the resonator must be transported using an iSWAP gate into the designated central qubit and be re-initialized there. The reset timescale is also somewhat longer than when using a QCR. 

\section{Results and discussion}
\label{section:results}

\begin{figure*}

  \includegraphics[width=0.99\linewidth]{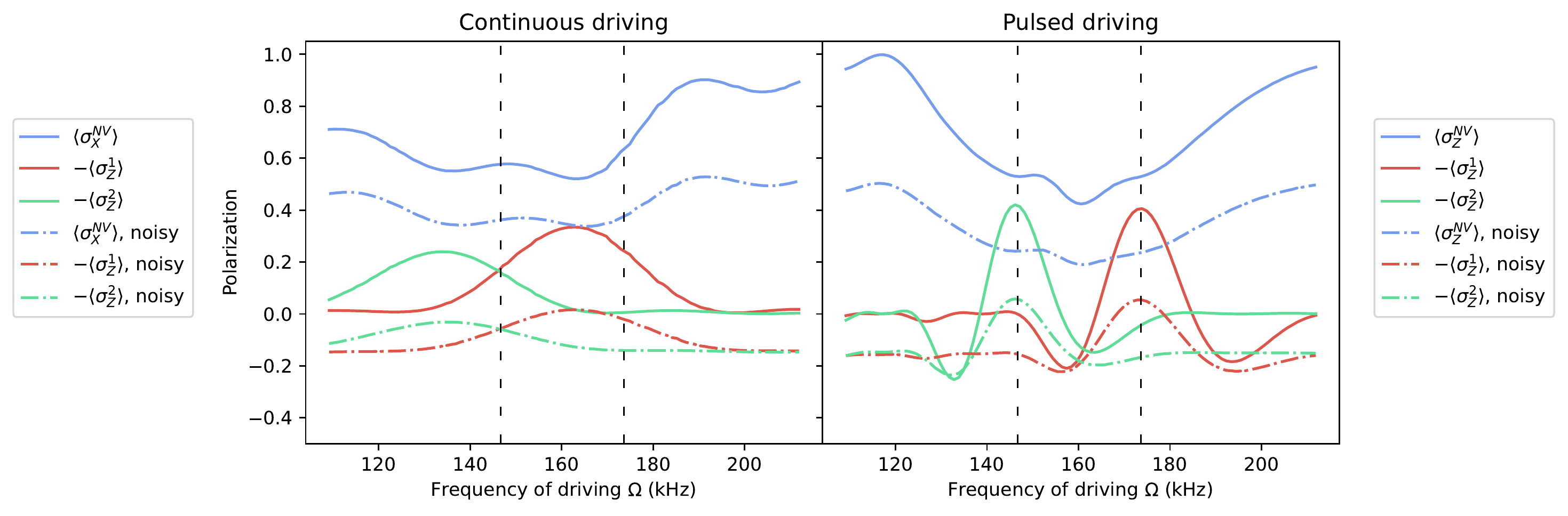}%
  \label{fig:sim}%

	\caption{Polarization transfer from one NV center to two interacting nuclei for a simulation time $t_f = 30 \, \mu $s and a single cycle with $s=32$ Trotter steps. The relevant observables to represent polarization are those included in the legends. (Left) with continuous driving and (right) with pulsed driving. Both plots depict with solid lines an exact simulation of the nanoscale NMR system including its usual imperfections, namely a detuning of $\delta_1 = 120$ kHz and a fluctuating microwave drive amplitude which follows an OU process with correlation time $\tau = 500 \, \mu$s and diffusion constant $c = 4\cdot 10^{-7} \, \mu$s$^{-1}$ (see Eq. (\ref{eq:OU})). The dotted lines show a simulation additionally including QPU noise and errors (as defined in Sec.~\ref{subsec:noise_and_errors}). The dashed vertical black lines indicate the expected resonance frequencies of the nuclei, i.e. where the peaks should be centered in the absence of the detuning $\delta_1$. The noisy QPU has $\varepsilon _{\textrm{SQG}} = 10^{-4}$, $\varepsilon _{\textrm{TQG}} = 2\cdot 10^{-3}$ and amplitude damping and pure dephasing with  $T_1 = 60\, \mu$s and $T_2 = 60 \, \mu $s with gate durations of single- and two-qubit gates $\tau_{\textrm{SQG}}=60 \, $ns and $\tau_{\textrm{TQG}}=27 \, $ns respectively. The parameters are chosen to demonstrate the performance of a state-of-the-art superconducting QPU. The characteristic effect of the detuning is to shift the curves to the left in frequency domain, as can be seen in (left). In (right) this effect is compensated by the pulsed driving, which refocuses detuning errors. The amplitude damping affecting the QPU shifts down the expectation value of all observables, while the (depolarizing) gate errors decrease the polarization transfer efficiency by reducing the visibility of the peaks.}
	\label{fig:simulation_results}
\end{figure*}

In this section we discuss the two main results of the paper: namely the predicted performance of our proposed quantum algorithm on a regular noisy QPU, as well as the performance increase obtained with our proposed Co-Design QPU. To this aim, we will focus on the polarizations of the NV center and nuclear spins, that are relevant quantities of the problem and straightforward to measure in a quantum computer.

In Fig.~\ref{fig:simulation_results} we compare the frequency response of the polarization transfer process on two different simulated devices: a QPU with realistic noise parameters, and an ideal noiseless QPU. We consider one NV center, two interacting nuclei and different driving frequencies for both continuous and pulsed driving schemes. In the simulation we ignore errors in the preparation of the fully-mixed state of the qubits representing the nuclei. The blue curves show the remaining polarization in the NV center after one cycle of initialization and time evolution, while the red and the green curves correspond to the nuclear polarizations at the end of the cycle. For each nucleus there appears a resonance frequency in the system, for which the polarization transfer is optimal for said nucleus, depicted in the figure by the peaks of the curves.

Both simulations include the effects of the most common imperfections in nanoscale NMR systems, i.e. energy detunings and Rabi frequency fluctuations discussed in Sec.~\ref{sec:hyperpolarization}. The simulation of the quantum algorithm additionally includes noise and gate errors present in the QPU. It is notable that the noise affects the height and shape of the peaks more than their location.

The system imperfections include a detuning of $120$ kHz of the NV center from the zero-field splitting that shifts the peaks in Fig.~\ref{fig:simulation_results} (left) to frequencies lower than their predicted Larmor frequencies (dotted vertical black lines). Fig.~\ref{fig:simulation_results} (right) shows how the pulsed-driving scheme XY8~\cite{maudsley1986modified,gullion1990new} acts as a robust dynamical decoupling sequence, eliminating such frequency shifts both in the ideal and noisy simulations. 

Regarding the QPU noise and errors, the amplitude damping channel causes an overall shift down of all polarizations at all driving frequencies.
Dephasing noise and gate errors (as modelled by depolarizing noise) cause the curves in Fig.~\ref{fig:simulation_results} to flatten and lose contrast. While we have discussed how the product of gate errors is minimized by reducing the SWAP overhead through Co-Design hardware, the loss of contrast can also be addressed through error mitigation techniques such as zero-noise extrapolation~\cite{endo2018practical,cai2021multi,krebsbach2022}. Dephasing can also be reduced through dynamical decoupling techniques~\cite{Krantz2019}, thus extending the system coherence and increasing the effective $T_2$ time. The simulations presented in Fig.~\ref{fig:simulation_results} include the decoherence times and gate fidelities that can be achieved with the hardware in Sec.~\ref{sec:co-design_HW}. This implies an overestimation of  the actual errors in the simulation, since the gate fidelities already include some decoherence.

\begin{figure}[h!btp]
    \subfloat[\label{subfig:comparison_codesign_signal_to_noise_error}]{%
        \includegraphics[width=0.99\linewidth]{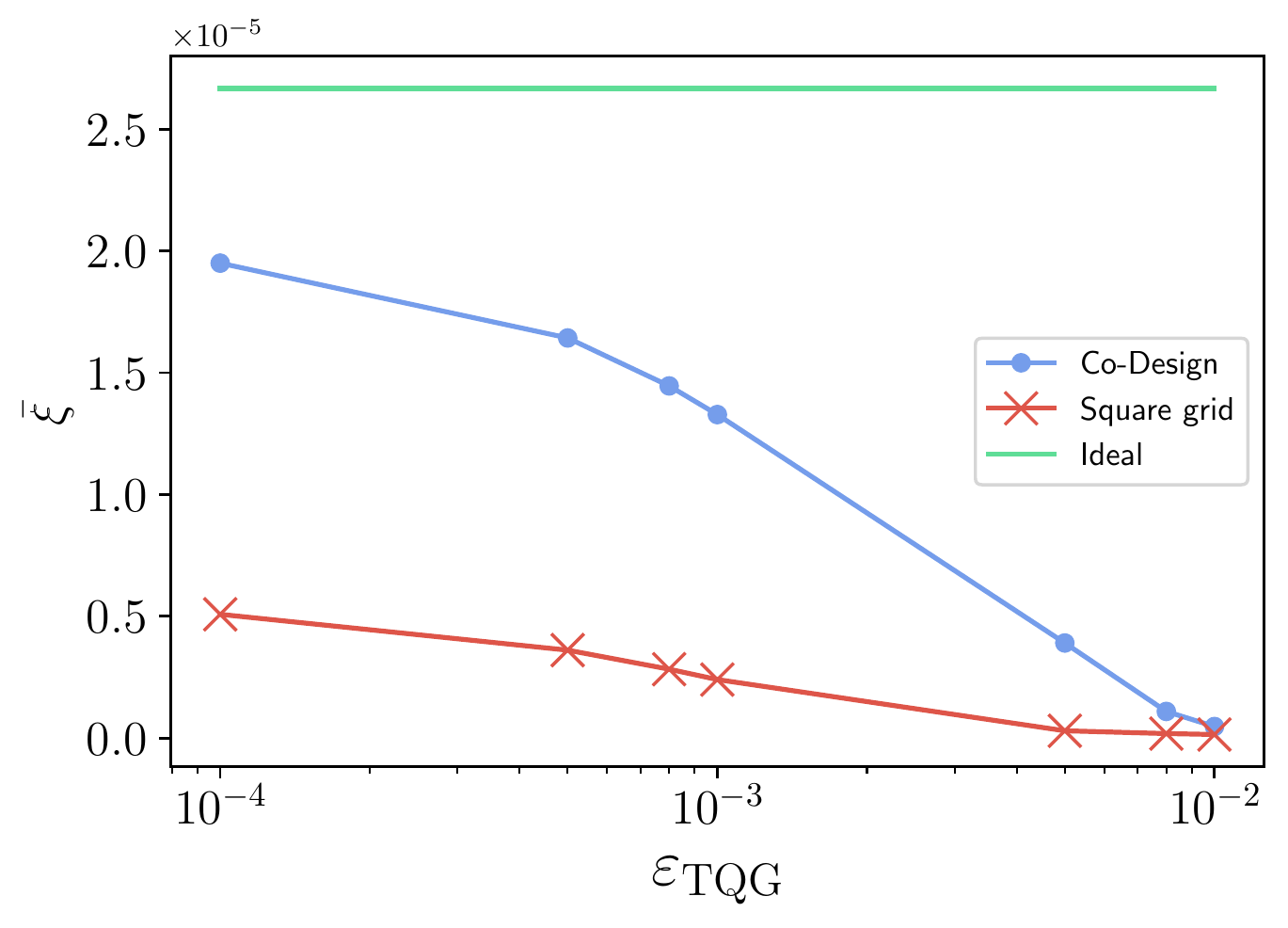}
    }\hfill
    \subfloat[\label{subfig:comparison_codesign_peak}]{%
        \includegraphics[width=0.99\linewidth]{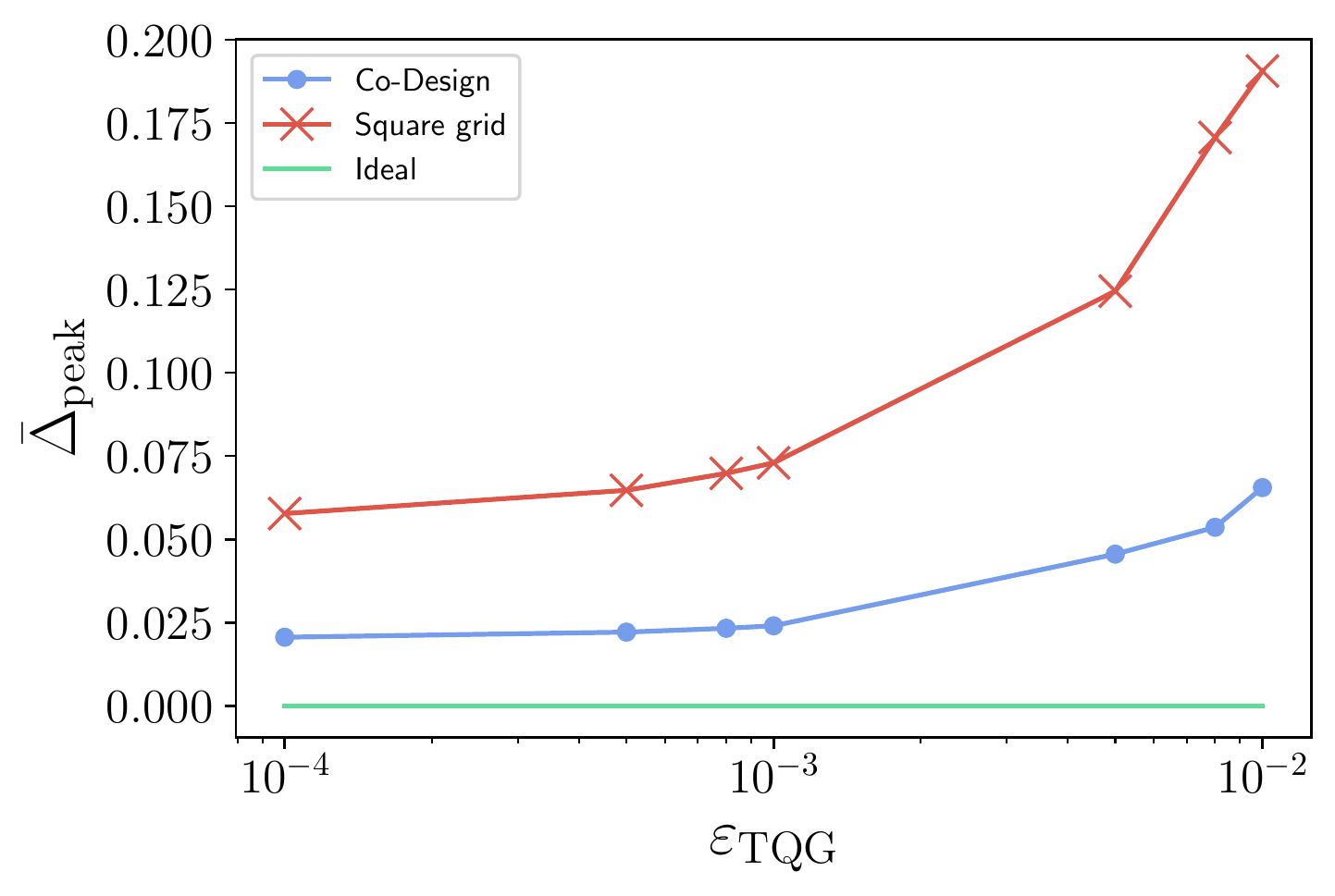} 
    }\hfill
    \caption{Performance gain from Co-Design: a comparison between a Co-Design star-architecture against a square grid, taking as reference an ideal simulation without QPU noise. The comparison highlights the negative effect that the SWAP gates on the square grid have on extracting relevant information from the simulation. We consider two quantities: in subplot (a) the ratio  $\bar{\xi}$  between the height and the width of the polarization peaks, and in subplot (b) the estimation error of the polarization peak center $\bar{\Delta}_{\textrm{peak}}$, where the bars denote an average over five nuclei for each noise level. The simulations were performed with the same parameters as in Fig.~\ref{fig:simulation_results}, except for the number of qubits, which has been increased from 2 to 5.}\label{fig:comparison_codesign_}
\end{figure}

To quantify the advantage of our Co-Design processor, Fig.~\ref{fig:comparison_codesign_} shows how the reduction in TQGs improves our ability to extract relevant information from the simulation. The figure compares the star-architecture chip to qubits connected on a square grid simulating a six qubit system with one NV center and five non-interacting nuclei. On the two chips we use SWAP patterns according to the schemes discussed in Sec.~\ref{subsection:layout}.

First, Fig.~\ref{subfig:comparison_codesign_signal_to_noise_error} shows the average height-to-width ratio $\bar{\xi}$ of the nuclear polarization peaks obtained with star and square grid topologies with respect to an ideal error-free simulation. It serves as an indicator of how much the QPU noise degrades the simulation for each case. The ratio $\bar{\xi}$ is computed by fitting a Gaussian function on each peak, and computing:
\begin{equation}
\bar{\xi} = \Big{\langle}\frac{h}{\sigma}\Big{\rangle},
\end{equation}
where $h$ is the height and $\sigma$ the variance of the fitted Gaussian function, averaged over the five nuclei. 

The curves for both topologies must coincide at $\bar{\xi} = 0$ for a maximal-error device, and at $\bar{\xi} = \bar{\xi}_{\textrm{ideal}}$ for an error-free quantum computer, since for a maximal-error device the output is pure noise and for an error-free quantum computer the number of SWAPs is irrelevant to the precision. For NISQ devices in between these limits, a performance difference between the architectures is observed. For systems with more nuclei and NV centers, the differences between topologies start to appear at lower errors, since the number of total operations grows. This shows how the QPU topology is of great importance for the computational precision of NISQ devices, while for fault-tolerant quantum computers the precision is unaffected by the topology. 

Second, Fig.~\ref{subfig:comparison_codesign_peak} shows the average relative error in the central frequency of the NMR peaks:
\begin{equation}
\bar{\Delta}_{\textrm{peak}} = \Big{\langle}\Big|\frac{\omega_{\textrm{noisy}}-\omega_{\textrm{ideal}}}{\omega_{\textrm{ideal}}}\Big|\Big{\rangle},
\end{equation}
where $\omega_{\textrm{noisy}}$ and $\omega_{\textrm{ideal}}$ are the peak-center frequencies extracted from the Gaussian fittings for the noisy and ideal cases, respectively. The peak centers correspond to driving frequencies that efficiently transfer polarization to different parts of the diamond lattice. 

With the quantum simulation we can individually identify the nuclear resonance peaks by directly  measuring the polarization of each qubit. This could enable exploration of how the polarization diffuses in the lattice with single-nucleus precision. In contrast,  in a standard nanoscale NMR experiment, one typically only has only access to  the excitation loss of the NV ( and thus only to the average transmitted polarization). This demonstrates the advantage of simulating the system on a quantum computer, as a it provides access to the relevant microscopic details of the dynamics that are otherwise inaccessible.

The figures demonstrate that the Co-Design chip is able to detect the resonance frequencies and predict the peak heights better at all considered noise levels. The power of Co-Design is particularly evident in Fig.~\ref{subfig:comparison_codesign_peak}, where the square grid is shown to require two orders of magnitude lower noise levels to reach the same accuracy as the Co-Design chip.

\section{Conclusions and outlook}
\label{sec:conclusions}

We have presented a quantum algorithm to simulate a nanoscale NMR problem, namely a hyperpolarization protocol. We have simulated the proposed quantum algorithm with typical noise processes of a NISQ superconducting quantum computer with state-of-the-art parameters. We find that, despite considering a noisy QPU, our protocol still allows to identify the positions of the nuclear resonances (corresponding to the maximal polarizations) in the frequency domain, as well as the behavior in the vicinity of such resonant frequencies, thus enabling the exploration of optimized protocols and driving parameters to hyperpolarize the nuclear ensemble. 

Moreover, we have shown that a specific Co-Design architecture adapted to the problem provides  an advantage over general-purpose designs in the NISQ era, thanks to the reduction in two-qubit-gate count. Consequently, the adapted design reduces the necessary gate fidelities to solve practical problems in nanoscale NMR. This application-specific QPU consists of a central resonator, representing an NV center, coupled to a number of qubits representing the nuclei. The design can be scaled to more NV centers and a potentially large number of qubits around them. This is an example of a shortcut to quantum advantage. Adapting more NISQ-friendly algorithm alternatives, such as those listed in~\cite{bharti2021noisy}, adapted to the problem and to the Co-Design hardware can provide further shortcuts.

Our work opens interesting directions for further investigation, since a quantum processor able to efficiently simulate nanoscale-NMR scenarios with a large number of nuclear spins would have a great impact on NMR-based applications. Fast and reliable quantum simulations of interacting spin systems would improve the interpretability of zero- and low-field NMR where spin-spin interactions become dominant~\cite{seetharam2021digital}, and nanoscale-NMR systems where a quantum sensor is strongly coupled via dipole-dipole interactions to nuclear or electron spin clusters. A possible application of the latter is the estimation of inter-label distances (via, e.g., Bayesian analysis of the NV center response) in electronically labelled biomolecules~\cite{munuerajavaloy2021detection}. In this case, the numerical analysis of systems beyond two-electron spin labels in realistic conditions, including protein motion and decoherence channels, is already numerically challenging.

\section*{Acknowledgments} The authors would like to thank Caspar Ockeloen-Korppi, Alessandro Landra and Johannes Heinsoo for their help in developing the idea of the star-architecture chip, Jani Tuorila for his support in developing the gate theory, Amin Hosseinkhani and Tianhan Liu for reviewing the manuscript, and Henrikki M\"akynen and Hoang-Mai Nguyen for graphic design. J.C. additionally acknowledges the Ram\'on y Cajal program (RYC2018-025197-I). We further acknowledge support from Atos  with the Quantum Learning Machine (QLM). Finally, the authors acknowledge financial support to BMBF through the Q-Exa project No. FZK: 13N16062.

\appendix
\onecolumngrid
\section{Derivation of the system Hamiltonian}
\label{appendix:H_derivation}
The Hamiltonian in Eq.~\eqref{eq:hamiltonian1} can be derived from first principles. Let us first assume a model including only two $^{13}C$ nuclei and one NV center (Fig.~\ref{fig:interaction_scheme}) with dipole-dipole interactions. For simplicity we also consider the NVs to be aligned with the external magnetic field. In that case, the Hamiltonian of the system reads:

\begin{equation}
    H=D S_{z}^{2}-\gamma_{e} B_{z} S_{z}-\gamma_{c} B_{z}\left(I_{1}^{z}+I_{2}^{z}\right)+\sum_{k=1}^{2} \frac{\hbar \mu_{0} \gamma_{e} \gamma_{c}}{2\left|\vec{r}_{k}\right|^{3}}\left[\vec{S} \cdot \vec{I}_{k}-\frac{3\left(\vec{S} \cdot \vec{r}_{k}\right)\left(\vec{I}_{k} \cdot \vec{r}_{k}\right)}{\left|\vec{r}_{k}\right|^{2}}\right]+\frac{\hbar \mu_{0} \gamma_{c}^{2}}{2\left|\vec{r}_{1,2}\right|^{3}}\left[\vec{I}_{1} \cdot \vec{I}_{2}-\frac{3\left(\vec{I}_{1} \cdot \vec{r}_{1,2}\right)\left(\vec{I}_{2} \cdot \vec{r}_{1,2}\right)}{\left|\vec{r}_{1,2}\right|^{2}}\right],
\end{equation}
where $S_j$ is the $j$-th spin component of the NV center, $I^j_k$ the $j$-th spin component of nucleus $k$, $D$ is the zero-field splitting of the NV center, $\gamma_e$ and $\gamma_c$ are the gyromagnetic factors of the NV center and the nuclei respectively, $B_z$ is the external magnetic field, which is aligned with the symmetry axis of the NV center $\vec{r}_k$ is the relative position vector between the NV center and nucleus $k$ and $\vec{r}_{1,2}$ is the relative position vector between both nuclei. 

We go into an interaction picture with respect to $H_{0}=D S_{z}^{2}-\gamma_{e} B_{z} S_{z}$. The $\mathrm{NV}$-nuclei interaction term reads:
\begin{equation}
H_{\mathrm{NV}-\mathrm{N}}^{I}=\sum_{k=1}^{2} \frac{\hbar \mu_{0} \gamma_{e} \gamma_{c}}{2\left|\vec{r}_{k}\right|^{3}}\left\{\left[S_{z} I_{k}^{z}-\frac{3\left(S_{z} r_{k}^{z}\right)\left(\vec{I}_{k} \cdot \vec{r}_{k}\right)}{\left|\vec{r}_{k}\right|^{2}}\right]+U_{0}^{\dagger}\left[S_{x} I_{k}^{x}+S_{y} I_{k}^{y}-\frac{3\left(S_{x} r_{k}^{x}+S_{y} r_{k}^{y}\right)\left(\vec{I}_{k} \cdot \vec{r}_{k}\right)}{\left|\vec{r}_{k}\right|^{2}}\right] U_{0}\right\},
\end{equation}
where we split the expression in commuting and non-commuting operators. The non-commuting operators pick a fast-rotating phase and can be neglected through the rotating-wave approximation. By performing an interaction-picture transformation with respect to $H_{0}=-\gamma_{c} B_{z}\left(I_{1}^{z}+I_{2}^{z}\right)=\omega\left(I_{1}^{z}+I_{2}^{z}\right)$, the nucleus-nucleus interaction term reads:
\begin{equation}
\begin{aligned}
H_{\mathrm{N}-\mathrm{N}}^{I}=&\frac{\hbar \mu_{0} \gamma_{c}^{2}}{2\left|\vec{r}_{1,2}\right|^{3}} U_{0}^{\dagger}\left[\vec{I}_{1} \cdot \vec{I}_{2}-\frac{3\left(\vec{I}_{1} \cdot \vec{r}_{1,2}\right)\left(\vec{I}_{2} \cdot \vec{r}_{1,2}\right)}{\left|\vec{r}_{1,2}\right|^{2}}\right] U_{0}= \\
&=\frac{\hbar \mu_{0} \gamma_{c}^{2}}{2\left|\vec{r}_{1,2}\right|^{3}}\left\{I_{1}^{z} I_{2}^{z}+\frac{1}{2}\left(I_{1}^{+} I_{2}^{-}+I_{1}^{-} I_{2}^{+}\right) - \right. \\
& \left. -\frac{3\left[I_{1}^{+} e^{i \omega t}\left(r_{1,2}^{x}-i r_{1,2}^{y}\right)+I_{1}^{-} e^{-i \omega t}\left(r_{1,2}^{x}+i r_{1,2}^{y}\right)\right]\left[I_{2}^{+} e^{i \omega t}\left(r_{1,2}^{x}-i r_{1,2}^{y}\right)+I_{2}^{-} e^{-i \omega t}\left(r_{1,2}^{x}+i r_{1,2}^{y}\right)\right]}{4\left|\vec{r}_{1,2}\right|^{2}}\right\},
\end{aligned}
\end{equation}
with $I_{k}^{\pm}=I_{k}^{x} \pm i I_{k}^{y}$. Applying again the rotating-wave approximation and undoing the interaction picture we finally arrive at:
\begin{equation}
H_{I}=-\gamma_{c} B_{z}\left(I_{1}^{z}+I_{2}^{z}\right)+S_{z}\left(\vec{A}_{1} \cdot \vec{I}_{1}+\vec{A}_{2} \cdot \vec{I}_{2}\right)+g_{1,2}\left[I_{1}^{z} I_{2}^{z}-\frac{1}{4}\left(I_{1}^{+} I_{2}^{-}+I_{1}^{-} I_{2}^{+}\right)\right],
\label{eq::hamilt_inter_appendix}
\end{equation}
with $\vec{A}_{k}=\frac{\hbar \mu_{0} \gamma_{e} \gamma_{c}}{2\left|\overrightarrow{r_{k}}\right|^{3}}\left[\hat{z}-\frac{3\left(\hat{z} \cdot \vec{r}_{k}\right) \vec{r}_{k}}{\left|\vec{r}_{k}\right|^{2}}\right]$, and $g_{1,2}=\frac{\hbar \mu_{0} \gamma_{c}^{2}}{2\left|\vec{r}_{1,2}\right|^{3}}\left[1-3\left(\frac{r_{1,2}^{z}}{\left|\vec{r}_{1,2}\right|}\right)^{2}\right]$.

We rewrite $S_{z}$ in the $\{|0\rangle,|1\rangle\}$ subspace by dropping out the $|-1\rangle$ energy state as it will not participate in the dynamics. Leakage to that state would not be a problem because of the energy difference between states $|0\rangle,|1\rangle$ and $|0\rangle,|-1\rangle$. Then by using that $|1\rangle\langle 1|=\frac{\mathbb{1}-\sigma^{z}}{2}$ to we get:

\begin{equation}
H_{I}=-\vec{\omega}^{c}_{1} \cdot \vec{I}_{1}-\vec{\omega}^{c}_{2} \cdot \vec{I}_{2}+\frac{\sigma_{z}}{2}\left(\vec{A}_{1} \cdot \vec{I}_{1}+\vec{A}_{2} \cdot \vec{I}_{2}\right)+g_{1,2}\left[I_{1}^{z} I_{2}^{z}-\frac{1}{4}\left(I_{1}^{+} I_{2}^{-}+I_{1}^{-} I_{2}^{+}\right)\right],
\label{eq:interaction_hamiltonian}
\end{equation}
where $\vec{\omega}^{c}_{k}=-\left(\frac{A_{k}^{x}}{2}, \frac{A_{j}^{y}}{2}, \frac{A_{j}^{z}}{2}-\gamma_{c} B_{z}\right)$ is the modified nuclear Larmor term due to the presence of the NV center. 

Generalizing equation~\eqref{eq:interaction_hamiltonian} to $M$ NV centers and $N$ nuclei, including the detuning of the NV centers and adding the microwave driving term we obtain precisely the Hamiltonian in Eq.~\eqref{eq:hamiltonian1}.

\section{\label{HH_sequence} Hyperpolarization sequences}
\subsection{Hartmann-Hahn sequence}
Here we explain the dynamics induced by the continuous driving on the hyperpolarization protocol. To illustrate the mechanism, we consider a system including a single NV center and a single nucleus. The corresponding Hamiltonian, now including the driving term, reads:

\begin{equation}
H=D S_{z}^{2}-\gamma_{e} B_{z} S_{z}-\gamma_{c} B_{z}I_{z}+S_{z}\vec{A}\cdot \vec{I} +S_{x} \sqrt{2} \, \Omega \cos (\omega t-\phi),
\end{equation}

In the interaction picture with respect to $D S_{z}^{2}-\gamma_{e} B_{z} S_{z}$ we obtain:

\begin{equation}
H_{I}=-\gamma_{n} B_{z} I_{z}+S_{z} \vec{A} \cdot \vec{I}+\frac{\Omega}{2} \left( e^{i p_{+} t}|1\rangle\langle 0|+e^{i p_{-} t}|-1\rangle\langle 0|+\textrm{H.c.} \right) \left[e^{i(\omega t-\phi)}+e^{-i(\omega t-\phi)}\right],
\end{equation}
where $p_{+/-}=D \pm\left|\gamma_{e}\right| B_{z}$. Choosing the resonance condition $\omega=p_{+}$ and applying the rotating-wave approximation we get:
\begin{equation}
H_{I}=-\gamma_{n} B_{z} I_{z}+S_{z} \vec{A} \cdot \vec{I}+\frac{\Omega}{2}\left(e^{i \phi}|1\rangle\langle 0|+e^{-i \phi}|0\rangle\langle 1|\right).
\end{equation}
Finally we can use the identity $|1\rangle\langle 1|=\frac{\mathbb{1}-\sigma^{z}}{2}$ and the fact that there will be no transitions to the $|-1\rangle$ because of energy differences:

\begin{equation}\label{seq:simphamiltonian}
H_I = -\vec{\omega}^c\cdot\vec{I} - \frac{\sigma_z}{2}\vec{A}\cdot\vec{I} + \frac{\Omega}{2}\sigma^\phi,
\end{equation}
where $\sigma^\phi=e^{-i\phi} |1\rangle\langle 0|+e^{i\phi} |0\rangle\langle 1| = e^{-i\phi}\sigma^{-}+e^{i\phi} \sigma^{+}$ and $\vec{\omega}_{n}=-\left(\frac{A_{x}}{2}, \frac{A_{y}}{2}, \frac{A_{z}}{2}-\gamma_{n} B_{z}\right)$. More details about the different terms were discussed in the main text, in section~\ref{sec:hyperpolarization}. Choosing $\phi = 0$ and further moving to an interaction picture with respect to the terms $-\vec{\omega}^c\cdot\vec{I} + \frac{\Omega}{2}\sigma_x$ we obtain:
\begin{equation}
H_I =\frac{e^{i\frac{\Omega}{2}\sigma_x t}\sigma_{z}e^{-i\frac{\Omega}{2}\sigma_x t}}{2}e^{-i\vec{\omega}^c\cdot \vec{I}t}\vec{A}\cdot\vec{I}e^{i\vec{\omega}^c\cdot \vec{I}t}.
\end{equation}
We choose now $\Omega = |\vec{\omega}^c|$, leading to the so-called Hartmann-Hahn double-resonance condition. Applying the identity $e^{i \vec{I}\cdot \hat{l}\phi} \vec{I}\cdot\vec{b} e^{-i \vec{I}\cdot \hat{l}\phi} = \vec{I}\left[(\vec{b}-(\vec{b}\cdot\hat{l})\hat{l})\cos{\phi}-\hat{l}\times\vec{b}\sin{\phi}+(\vec{b}\cdot\hat{l})\hat{l}\right]$ and the rotating-wave approximation to remove time-dependent terms, we get the flip-flop Hamiltonian:
\begin{equation}
H_I= \frac{A^\perp}{4}\left(|+\rangle\langle-|I^++|-\rangle\langle+|I^-\right),
\label{eq:flipflopcont}
\end{equation}
with $A^\perp = \left|\vec{A}_x^\perp\right| =  \left|\vec{A}-\left(\vec{A}\cdot\hat{\omega}^c\right)\hat{\omega}^c\right|$ and the nuclear coordinates changed so that $\hat{x} = \hat{A}_x^\perp$ and $\hat{z} = \hat{A}_z^\parallel$ with $\vec{A}_z^\parallel = (\vec{A}\cdot\hat{\omega}^c)\hat{\omega}^c$.

\subsection{Pulsed sequence}
\label{subsec:pulsed}

Now we consider the pulsed case, represented by the driving term $H_{\textrm{dr}} = \frac{\Omega(t)}{2}\sigma^\phi$ where $\Omega(t)$ is a train of $\pi$-pulses. The Hamiltonian is already expressed in the interaction picture from Eq.~\eqref{seq:simphamiltonian}. From there, we further move into a rotating frame with respect to the driving term. The corresponding unitary transformation is $U_0 = (-i \sigma^\phi)^k$ for the time interval between pulses $k$ and $k+1$. This leads to:
\begin{equation}
H_I =  -\vec{\omega}^c\cdot\vec{I} + F(t) \frac{\sigma_z}{2}\vec{A}\cdot\vec{I},
\label{eq:filter}
\end{equation}
where $F(t)$ is the so-called filter function, with value $+1$ when $k$ is even, and $-1$ when $k$ is odd, representing the sign of the operator $\sigma_z$, flipped by the action of each pulse.

It is necessary to apply two different patterns of pulses. The "symmetric case", meaning an evenly-distributed sequence of pulses for which the filter function is even and can be expanded in Fourier series of cosines as: 
\begin{equation}
F(t) =  \sum_{n = 1}^\infty f_n \cos\left(\frac{2\pi n}{T} t\right),
\end{equation}
with $f_n = 0$ when $n$ is even and $f_n = -\frac{4}{\pi n}$ when $n$ is odd, if the pulses are distributed such that the interpulse spacing is constant. We choose the resonance condition $T = \frac{2\pi n}{|\vec{\omega}^c|}$, where $n$ is the harmonic number. This is the same resonance condition that we introduced in section \ref{sec:hyperpolarization}, but here it is formulated with the period $T$ that appears in the Fourier expansion, instead of with the interpulse spacing $\tau = \frac{T}{2}$ from before. Going to an interaction picture with respect to $ -\vec{\omega}^c\cdot\vec{I}$ and repeating the procedure we used above in the Hartmann-Hahn case, we get:
\begin{equation}
H_I = \alpha A^\perp\sigma_zI_x,
\end{equation}
where $\alpha = \frac{f_n}{4}$.

With the second pattern of pulses, called the "asymmetric case", we apply an oddly-distributed sequence of pulses for which the filter function is odd and can be expanded in a Fourier series of sines. Note that this sequence of pulses is identical to the even sequence but shifted by a $\pi/2$ phase. An analogous derivation gives:

\begin{equation}
H_I = \beta A^\perp\sigma_zI_y,
\end{equation}
with $\beta = \frac{g_m}{4}$ and $g_m$ coming from the Fourier expansion of sines, analogously to $f_n$.

Combining these two patterns one can generate an effective Hamiltonian of the form:

\begin{equation}
H_I = \alpha A^\perp\sigma_zI_x + \beta A^\perp\sigma_zI_y,
\end{equation}
which can be transformed with simple rotations on the qubit representing the NV into:
\begin{equation}
H_I = \alpha A^\perp\sigma_xI_x + \beta A^\perp\sigma_yI_y,
\label{hamilt_pulsed}
\end{equation}
and this is equivalent to an interaction-exchange flip-flop Hamiltonian, similar to the one for the continuous-driving case (\ref{eq:flipflopcont}). A more detailed description of this whole process, including the expressions of the Fourier coefficients $f_n$ and $g_m$ can be found in reference~\cite{MunueraJavaloy2021}.

\begin{figure*}[h!]
    \subfloat[\label{subfig:algorithm_sketch_pulsed}]{%
        \includegraphics[width=0.99\linewidth]{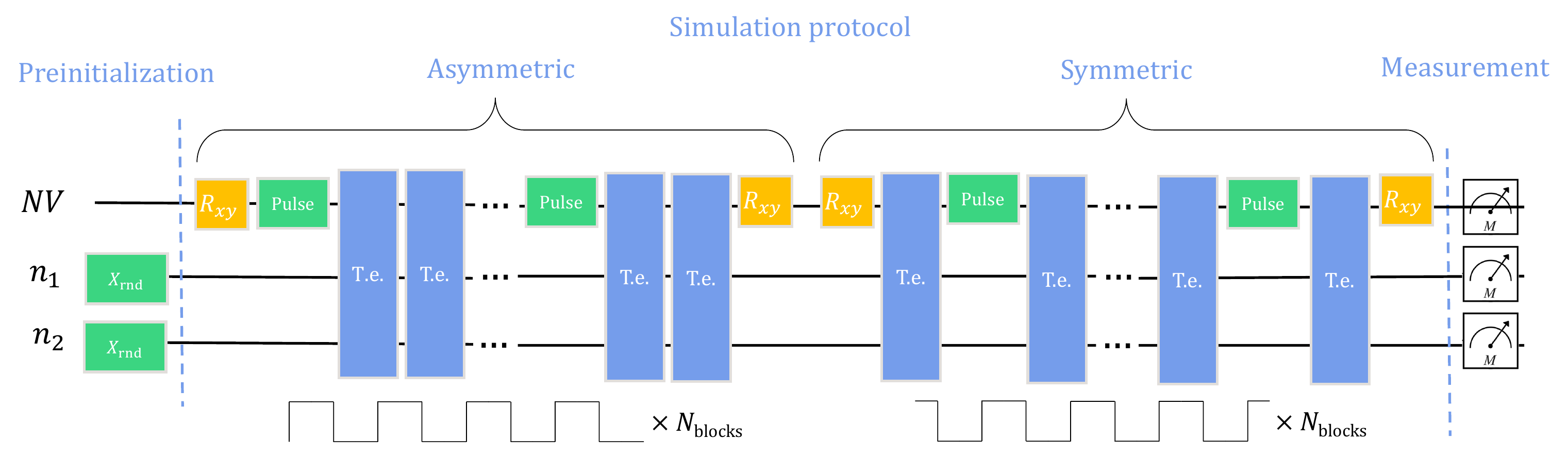}
    }\hfill
    \centering
    \subfloat[\label{subfig:one_trotter_step_pulsed}]{%
        \includegraphics[width=0.99\linewidth]{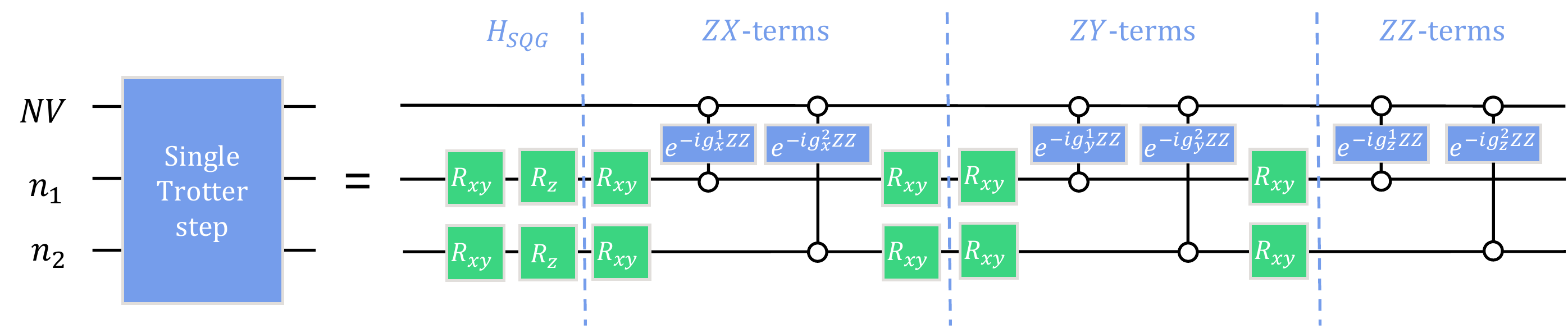} 
    }\hfill
        \caption{(a) Sketch of one cycle of the simulation algorithm for one NV center and two nuclei, with pulsed driving. Compare with Fig. \ref{fig:circuit_comparison_graphics} (where several cycles were depicted). The $R_{xy}$ gates highlighted in yellow are the rotations on the qubit representing the NV mentioned just before equation (\ref{hamilt_pulsed}). The term T.e. stands for Trotterized evolution and represents half of the free evolution of the system in between pulses, and can be devided into one or more single Trotter steps. The asymmetric and symmetric sequences of pulses are the ones discussed in Appendix \ref{subsec:pulsed}. The schematic drawings below the circuit in the form of square waves depict the modulation of the filter function (eq. (\ref{eq:filter})) under the two different pulse patterns. We choose the pulses to be either $X$ or $Y$ gates acting on the qubit representing the NV, following the pattern \textit{XYXYYXYX}, which can be repeated $N_{\textrm{blocks}}$ times for a stronger signal amplification. However, in our simulations a single block was enough to see clear patterns of polarization transfer, such as the ones in the right plot of Fig. \ref{fig:simulation_results}. (b) Corresponding gate sequence of one Trotter step, as in Fig. \ref{fig:circuit_comparison_graphics}, but without the $H_{\textrm{dr}}$ inside, because in this case the driving is applied through the sequence of pulses.
    }\label{fig:circuit_comparison_graphics_pulsed}
\end{figure*}

In order to visualize the structure of the pulsed-driving case, we have included in Fig.~\ref{fig:circuit_comparison_graphics_pulsed} the circuit implementing all these terms on a quantum chip for the case of one NV center and two nuclei.

\newpage

\section{\label{section:other_sim}Randomized Trotter techniques}
As explained in section~\ref{sec:algorithm}, we chose Trotter expansion. Besides this, we can consider other simulation approaches such as the variational quantum simulator~\cite{yuan2019theory}, the quantum assisted simulator~\cite{bharti2021quantum}, numerical quantum circuit synthesis~\cite{younis2021qfast}, or a plethora of other quantum simulation algorithms aimed at NISQ devices~\cite{bharti2021noisy}.

In addition, other approaches like randomized Trotter have been recently shown to provide some advantage compared to standard Trotter expansions~\cite{Childs2019}. We also propose to use one randomized approach, qDRIFT~\cite{Campbell2019}, that consists of the following: instead of splitting the whole evolution operator $e^{-it_f\sum_j h_j H_j}$ into simpler terms as done in full Trotterization, the method applies a random selection of such terms to the quantum circuit. This random selection is based on the probability distribution given by the weight of each term $h_j H_j$. For a certain evolution time, this set of gates can approximate the whole evolution operator by statistically drifting the state of the circuit towards the deterministic final state.

The error bound for this method is given as~\cite{Campbell2019}:
\begin{equation}
\varepsilon^{\textrm{qDRIFT}}_{\textrm{sim}} \leq \frac{2\lambda^2 t_f^2}{N_{\textrm{terms}}},
\end{equation}where $\lambda = \sum_j h_j$ and $N_{\textrm{terms}}$ is the number of individual two-qubit evolution operators that are implemented. These evolution operators have the form $e^{-i\tau H_j}$, being $\tau$ a constant related to the relative weight $\frac{h_j}{\lambda}$ that the term $H_j$ has in the Hamiltonian.

The advantage of qDRIFT compared to Trotterization is particularly apparent when dealing with Hamiltonians with a large number of terms with small coefficients, simulated for short times. While in the standard Trotter case, every term has to be simulated for each step no matter how small its effect is, in qDRIFT this is not required. A more thorough analysis of errors in qDRIFT and gate counts can be found in~\cite{Chen2021}.

This method is particularly suitable to our problem, since the range of coefficients in the Hamiltonian of a real diamond is large due to the length scales involved.

In this case, with qDRIFT the terms with smaller coefficients  do not add a significant amount of gates as they would in conventional Trotterization approaches. 

We note that other adapted protocols such as SparSto~\cite{Ouyang2020} can further enhance the simulation of this type of systems. SparSto represents a compromise between Trotterization and qDRIFT, generally guaranteeing an equal or better performance than both of them. We will not go into detail on this method since Trotterization and qDRIFT are enough to illustrate the main ideas behind this work.

\section{\label{section:appendix_time_evol}Hamiltonian decomposition for Trotterized time evolution}

In order to simulate the dynamics generated by the Hamiltonian in Eq.~\eqref{eq:hamiltonian1} on a quantum computer using Trotterization, we first need to express it in a suitable way. To begin with, we split the Hamiltonian into two parts: 

\begin{equation}
   	H=H_{\textrm{SQG}}+H_{\textrm{TQG}},
   	\label{Hamilt_Paulis}
\end{equation}
which can be expressed in terms of qubit Pauli operators:

\begin{equation}
		H_{\textrm{SQG}} = \sum_{k=1}^{N} \Big [ \frac{A^x_k}{2}\frac{X_k}{2}+\frac{A^y_k}{2}\frac{Y_k}{2} + \Big (  \frac{A^z_k}{2}-\gamma_c B_z \Big )\frac{Z_k}{2} \Big ] +\sum_{j=1}^{M}  \delta_j Z_j,
\end{equation}

\begin{align}
	\begin{split}
		H_{\textrm{TQG}} &= \sum_{j=1}^{M} \sum_{k=1}^{N} \Big [ \frac{A^x_k}{2}\frac{X_k}{2}Z_j+\frac{A^y_k}{2}\frac{Y_k}{2}Z_j+\frac{A^z_k}{2}\frac{Z_k}{2}Z_j \Big ] +\\
		&+ \sum_{k'>k=1}^{N} \frac{g_{k'k}}{4} \Big [ Z_{k'} Z_k -\frac{1}{2} X_{k'} X_k - \frac{1}{2} Y_{k'} Y_k \Big ] +\\
		&+\sum_{j>j'}^M h_{j'j} \Big [Z_{j'} Z_j-X_{j'}X_j-Y_{j'}Y_j \Big ].
	\end{split}
\end{align}

Since in the rotating frame with the drive the Hamiltonian is time independent, the time-evolution operator is simply given by:
\begin{equation}
	U=e^{-i t_f H},
\end{equation}
where $t_f$ is the time for which the simulation runs.

The time-evolution operator is split into $s$ discrete steps through Trotter decomposition:
\begin{equation}
	U =e^{-it_f H} =e^{-it_f(H_{\textrm{SQG}}+H_{\textrm{TQG}})} \approx \left[ e^{-i\frac{t_f}{s} H_{\textrm{SQG}}}e^{-i\frac{t_f}{s}H_{\textrm{TQG}}}\right]^{s}+ \mathcal{O}\left( \left(\frac{t_f}{s} \right)^{2}\right).
\label{eq:first_splitting}
\end{equation}

The evolution operator associated with single-qubit gates in each Trotter step of equation~\eqref{eq:first_splitting} needs to be rewritten in terms of our native gate set. It is always possible to decompose any single-qubit unitary exactly, up to a global phase, into a sequence of three single-qubit rotations such as, for example, a rotation about the $y$-axis in between two rotations about the $z$-axis:
\begin{equation}
U_1 = R_{z}(\beta)R_{xy}(\pi/2, \gamma)R_{z}(\delta),
\label{eq:unitary}
\end{equation}
where the angles $\beta, \gamma,$ and $\delta$ need to be determined from the specific entries of the unitary in question to simulate the evolution of the $p^{_{\textrm{th}}}$ qubit:

\begin{equation}
    U_1^p = e^{-i\frac{t_f}{s}\left(\frac{A^x_p}{2}\frac{X_p}{2}+\frac{A^y_p}{2}\frac{Y_p}{2}+ \left(  \frac{A^z_p}{2}-\gamma_c B_z  \right)\frac{Z_p}{2}\right)}.
    \label{eq:uone}
\end{equation}

From now on, we will concentrate on the case of a single NV center, which will be encoded in qubit $0$. Then, the evolution operator associated to single-qubit gates for the NV center will be:
\begin{equation}
    U^0_1 = e^{-i\frac{t_f}{s}\delta_0 Z_0}.
    \label{eq:uonep}
\end{equation}

Matching the entries of the matrices corresponding to the unitaries on equations~\eqref{eq:uone} and~\eqref{eq:uonep} we get a system of equations for the angles $\beta, \gamma,$ and $\delta$ for each Trotter step $s$.

There are 3 (5) types of interaction terms of the form ${XZ,YZ,ZZ,\cdots}$ in $H_{\textrm{TQG}}$ without (with) internuclear interactions. Due to the native TQG being of only $ZZ$ interaction type (see Eq.~\eqref{ZZ_unitary}), local rotations need to be introduced for simulating the rest of the TQG terms. These are $R^{\sigma_i \rightarrow \sigma_j}_k$, which have the effect of converting the Pauli operator $\sigma_i$ into the Pauli operator $\sigma_j$ for qubit $k$.

After the Trotterization introduced in equation (\ref{eq:first_splitting}), the term $H_{\textrm{TQG}}$ corresponding to TQG contains some elements which do not commute with each other, and some of them which do commute with each other. We choose to split all terms in order to express the time-evolution operator in terms of the native gates that we assumed in section \ref{subsubsec:native_gates}. Only the elements that do not commute with each other contribute to the total Trotter error, which remains of the same order:

\begin{align} \label{eq:TQG_evolution}
	\begin{split}
e^{-i\frac{t_f}{s}H_{\textrm{TQG}}} \approx \, &e^{-i\frac{t_f}{s}\left(\sum_k \frac{A^x_k}{2}\frac{X_k}{2}Z_0\right)} e^{-i\frac{t_f}{s}\left(\sum_k \frac{A^y_k}{2}\frac{Y_k}{2}Z_0\right)} \\
&e^{-i\frac{t_f}{s}\left(\sum_k \frac{A^z_k}{2}\frac{Z_k}{2}Z_0\right)} e^{-i\frac{t_f}{s}\left(\sum_{k'>k} \frac{g_{k'k}}{4}Z_{k'}Z_k\right)}\\
&e^{i\frac{t_f}{s}\left(\sum_{k'>k} \frac{g_{k'k}}{8}X_{k'}X_k\right)}e^{i\frac{t_f}{s}\left(\sum_{k'>k} \frac{g_{k'k}}{8}Y_{k'}Y_k\right)}\\
&+ \mathcal{O}\left( \left(\frac{t_f}{s} \right)^{2}\right). 
	\end{split}
\end{align}

Finally, we observe that the operators $Z_kZ_0$ (and the rest of the TQG terms) commute with each other, so the exponentials can be further split without Trotterizing:

\begin{equation}
e^{-i\frac{t_f}{s}(\sum_k \frac{A^z_k}{2}\frac{Z_k}{2}Z_0)} = \Pi_k e^{-i\frac{t_f}{s}(\frac{A^z_k}{2}\frac{Z_k}{2}Z_0)}.
\end{equation}

The time-evolution operator implementing the continuous sinusoidal driving $\sigma^{\phi}$ is:

\begin{equation}
e^{-i\frac{t_f}{s}\frac{\Omega}{2}\sigma^\phi}=R_{xy}(-\phi,\theta = \Omega \frac{t_f}{s} ).
\end{equation}

The quantum algorithm for simulating the system under a pulsed-driving scheme is somewhat more involved than the continuous-driving case, due to the two different time-dependent processes involved in the Trotter decomposition: the free dynamics of the spins and the sequence of pulses.
The most crucial point to be aware of is the interplay between Trotter steps and interpulse spacing. 
The number of interpulse evolutions, i.e. number of pulses minus one, bounds from below the minimum number of Trotter steps for the simulation. Clearly, at least one Trotter step is needed for each interpulse evolution.

Taking this interplay into account, the most straightforward setup is to choose a frequency which will determine the spacing of the pulse sequence, and to identify each interpulse evolution with a single Trotter step. If the achieved precision is not high enough, more Trotter steps can be added for each interpulse evolution. Each $\pi$-pulse itself is simply implemented as an $X$- or $Y$-gate on the qubit representing the NV center. The OU-distributed Rabi frequency fluctuations present in nanoscale NMR systems are then simulated by over- and under-rotations of the $X$- and $Y$-gates.

\section{\label{section:rotational_opt}Rotational optimization}

In principle, we had a Hamiltonian with terms of the type $ZX$, $ZY$ and $ZZ$ for the case of no internuclear interactions. However, we can rotate the basis so the Hamiltonian loses the $ZX$ and $ZY$ terms, allowing to reduce the number of TQGs. To make up for this rotation, we need to introduce different constants \(\vec{A}^{\textrm{rot}}_i\) for the problem and rotate the vector state we obtain at the end before measuring it.  The rotations that we will consider are only one-qubit rotations on nuclei qubits and we are applying this just to the case with no internuclear interactions. Therefore, we can consider the effect of this rotation on only one qubit representing an arbitrary nucleus. We will exemplify this procedure using nucleus 1. If we want to obtain the mean value of \(\sigma_z\) acting on the nucleus:
\begin{equation}\label{eq:sigmaz1}
\begin{split}
\langle \sigma_z\rangle = \Tr \left(\rho(t_f) \sigma_z\right) = \Tr \left(U(0,t_f)\rho(0)U^\dagger(0,t_f)\sigma_z\right),
\end{split}
\end{equation}
where \(U(0,t_f)\) represents the evolution operator from \(t=0\) to \(t=t_f\). The density matrix $\rho(0)$ contains the state of the NV center (which is in the $|+\rangle$ state at $t=0$) and nucleus 1, i.e. $\rho(0)=|+\rangle \langle +|_{\textrm{NV}} \otimes \frac{\mathbb{1}_{1}}{2}$. Our intention is to obtain an expression of this mean value in terms of the rotated evolution operators and later, we will find the appropriate rotation to be perfomed. Then, taking into account that the trace is invariant under a rotation $R=\mathbb{1}_{\textrm{NV}} \otimes R_1$ we get:
\begin{equation}\label{eq:sigmaz2}
\langle \sigma_z\rangle = \Tr \left(RU(0,t_f)\rho(0)U^\dagger(0,t_f)\sigma_z R^\dagger \right) = \Tr \left(RU(0,t_f)R^\dagger R\rho(0)R^\dagger R U^\dagger(0,t_f)R^\dagger R\sigma_z R^\dagger \right).
\end{equation}

This can be expressed as:
\begin{equation}\label{eq:sigmaz3}
\begin{split}
\langle \sigma_z\rangle = \Tr \left(U_{\textrm{rot}}(0,t_f)\rho_{\textrm{rot}}(0)U_{\textrm{rot}}^\dagger(0,t_f) R\sigma_z R^\dagger \right).
\end{split}
\end{equation}

The density matrix of the nucleus is the identity. Thus, any rotation on nuclei qubits leaves the  density matrix unaffected, leading to:
\begin{equation}\label{eq:sigmaz4}
\begin{split}
\langle \sigma_z\rangle = \Tr \left(U_{\textrm{rot}}(0,t_f)\rho(0)U_{\textrm{rot}}^\dagger(0,t_f) R\sigma_z R^\dagger \right) .
\end{split}
\end{equation}
Then we need to rotate the system previous to the measurement. By using the invariance of the trace under cyclic permutations we get:
\begin{equation}\label{eq:sigmaz5}
\begin{split}
\langle \sigma_z\rangle = \Tr \left(R^\dagger U_{\textrm{rot}}(0,t_f)\rho(0)U_{\textrm{rot}}^\dagger(0,t_f) R\sigma_z \right),
\end{split}
\end{equation}
which is equivalent to introducing a counter-rotation in the circuit before measurement.

Now let us focus on the specific rotation we have to implement. Since the constants multiplying the Pauli matrices in the Hamiltonian are \(\frac{\vec{A_1}}{2}\) and \(\vec{\omega^c_1} = \frac{\vec{A_1}}{2}-\gamma_c B_z \vec{e_z}\) (for nucleus 1), we can  rotate the basis to obtain a representation in which the vectors have only $z$-component for \(\vec{A_1}\) and thus, $XZ$ and $YZ$ terms are removed. The vectors before and after the needed rotation can be seen in \hyperref[\detokenize{figures/fig-basis-rotation.png}]{Fig.\@~\ref{\detokenize{figures/fig-basis-rotation.png}}}.

\begin{figure}[t!]
\begin{center}
    \subfloat[\label{subfig:basis_rotation1.pdf}]{%
        \includegraphics[scale=0.7]{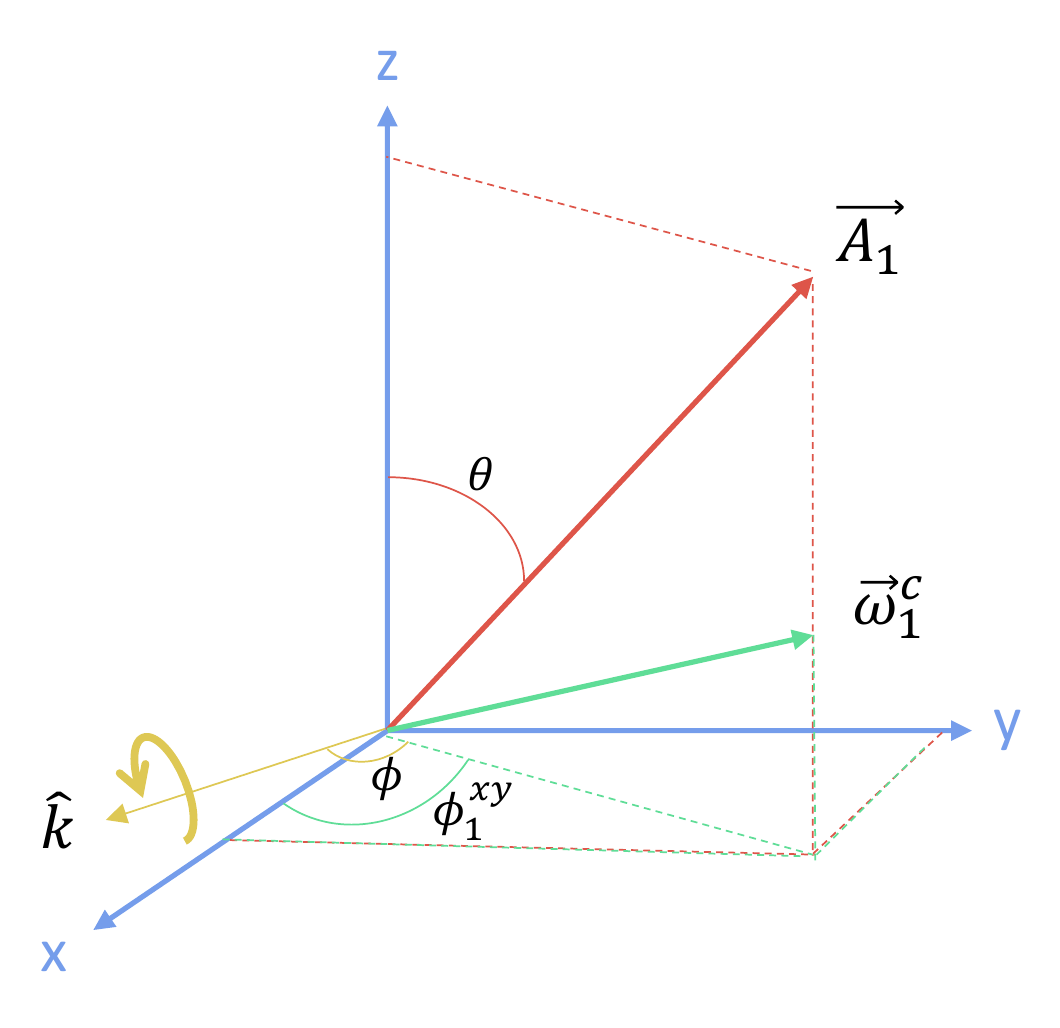}
    }\hfill
    \subfloat[\label{subfig:basis_rotation2.pdf}]{%
        \includegraphics[scale=0.7]{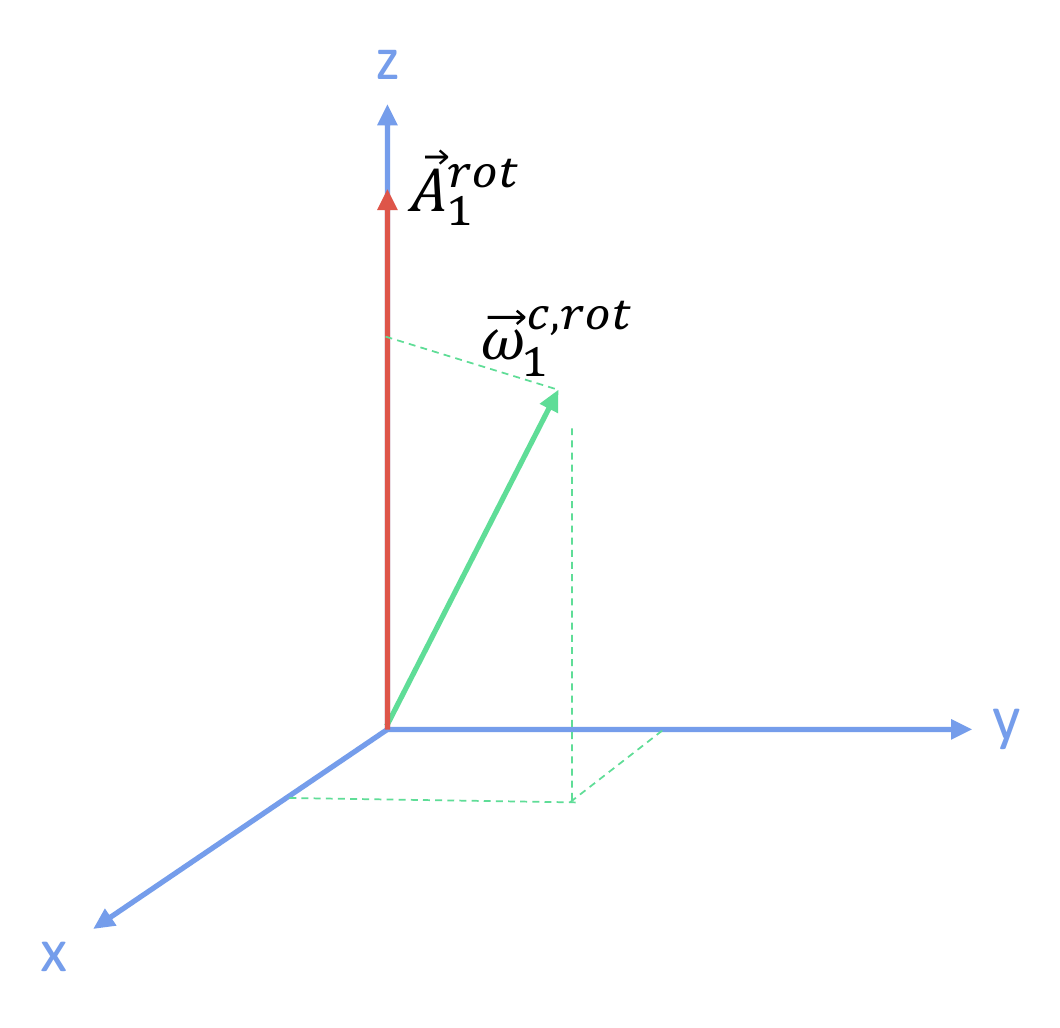} 
    }\hfill
    \caption{\(a)\) Coefficients vectors of the first qubit $\vec{A}_1$,$\vec{\omega}_1^c$ before the rotation, with projection over the three axis, \(b)\) coefficient vectors of the first qubit $\vec{A}^{rot}_1$,$\vec{\omega}_1^{c,\textrm{rot}}$ after the rotation, being \(\vec{A}^{\textrm{rot}}_1\) in the Z-axis.}\label{\detokenize{figures/fig-basis-rotation.png}}
\end{center}
\end{figure}

To compute the new vectors (and thus the new coefficients for the gates of our algorithm), we can use Rodrigues' rotation formula to rotate a vector $\vec{v}$ an angle $\theta$ around a unitary axis $\hat{k}$:
\begin{equation}
\vec{v}_{rot} = \vec{v} \cos\theta + (\hat{k} \times \vec{v}) \sin \theta + \hat{k}(\hat{k}\cdot \vec{v})(1- \cos\theta),
\end{equation}
being in our case, $\theta = \arccos{(A^z_1/|\vec{A}_1|)}$ and $\hat{k} = (\cos(\phi),\sin(\phi),0)$, with $\phi = -\frac{\pi}{2}+\phi_{xy}= -\frac{\pi}{2}+\arctan{(A^y_1/A^x_1)}$.

For implementing the counter-rotation of this in the quantum circuit, we use:
\begin{equation}
R_1^{\dagger} = e^{i\frac{\theta}{2}(\cos(\phi)X-\sin(\phi)Y)}.
\end{equation}

\section{\label{sec:routing}SWAP routing}

Our qubit routing method consists of mapping the square grid to a linear chain with qubits labeled from 0 to $n$. Then, in the simplified case of no internuclear interactions, the optimal SWAP method for the one-to-all interaction case on a linear chain can be used. For a single NV center the protocol goes as follows:
\begin{enumerate}
\item Initialize the state of the NV center in the second qubit;
\item Perform interactions with the first and third qubits;
\item SWAP the NV center qubit to the right;
\item Perform interaction with right qubit;
\item Repeat steps 3-4 until all interactions have been achieved.
\end{enumerate}
The pattern is seen in Fig.~\ref{fig:swap_patterns}a denoted by the intense blue arrows.
With internuclear interactions we need to perform a swap pattern that enables all-to-all interactions. The so-called odd-even mapping in Fig \ref{fig:swap_square} is an efficient one~\cite{Cowtan2019} represented by green arrows in Fig.~\ref{fig:swap_patterns}a. This consists of swapping first all the even qubits with their right neighbors and then swapping all the odd qubits with their right neighbors. This way, we will obtain all-to-all interactions with $\frac{1}{2}(n-1)(n-2)$ SWAP gates and a total TQG depth of $6n$. A summary of the TQG counts is shown in Table~\ref{tab:gate_count}.

\begin{figure}[h!btp]
    \subfloat[\label{subfig:swap_comp_with_numerical1}]{%
        \includegraphics[width=0.49\linewidth]{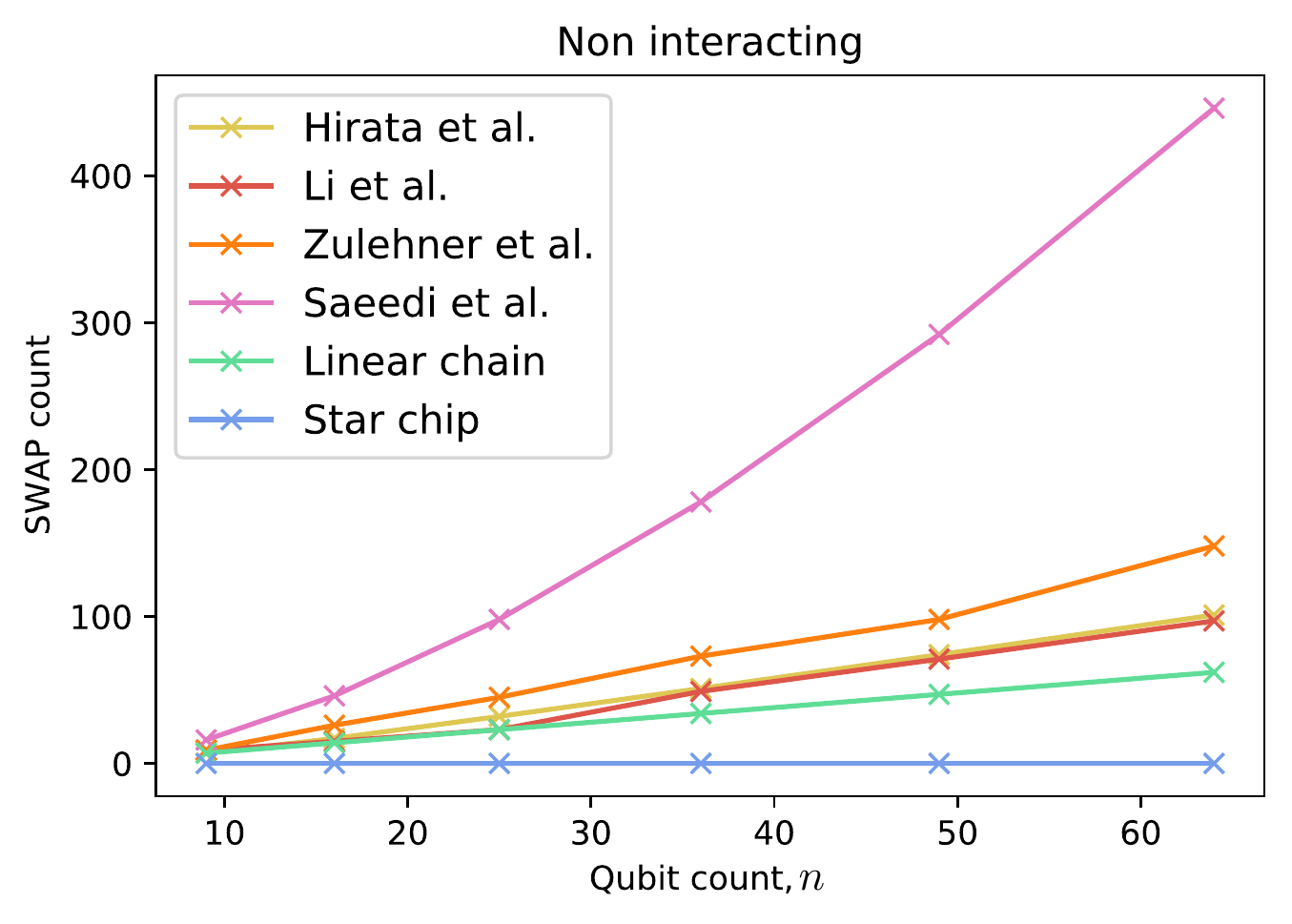}
    }\hfill
    \subfloat[\label{subfig:swap_comp_with_numerical2}]{%
        \includegraphics[width=0.49\linewidth]{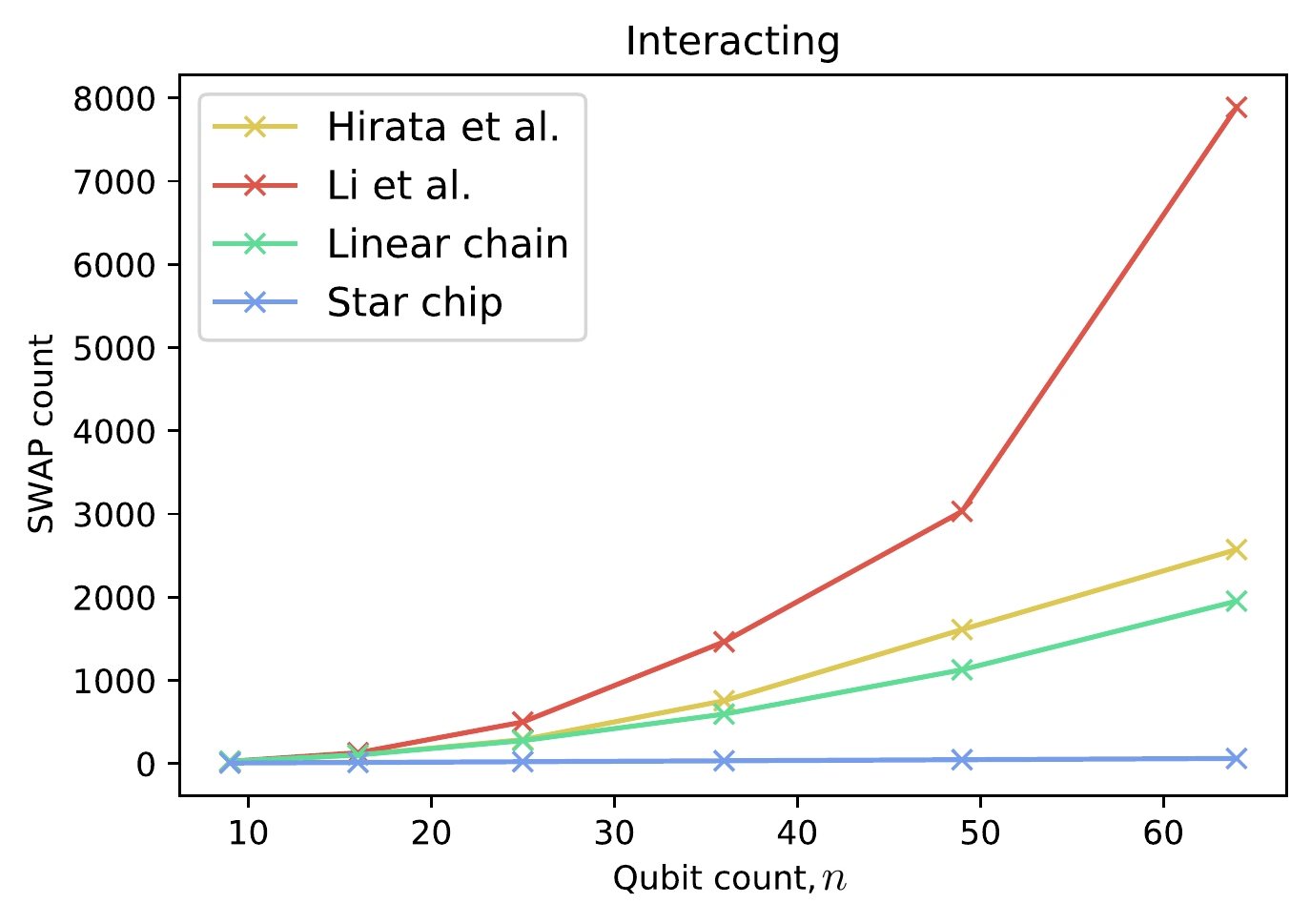} 
    }\hfill
    \caption{a) Comparison of the required number of SWAPs for simulating the proposed system with no internuclear interactions for each Trotter step. Numerical approaches from references are applied to a square grid. b) Equivalent comparison with internuclear interactions. Zulehner et al. and Saeedi et al. do not improve the linear chain approach for few qubits and are intractable for larger numbers of qubits and thus are not displayed.}\label{fig:swap_comp_with_numerical}
\end{figure}

To motivate the creation of a chip with a star topology and the use of an alternative linearized SWAP routing for a square grid instead of standard numerical approaches, a comparison between all the cases is provided in Fig. \ref{fig:swap_comp_with_numerical}. A reduction in the number of SWAPs can be noticed for both the linear chain approach and the star-topology chip against standard numerical approaches for a square grid.

\begin{center}
\begin{table}[]
\def\arraystretch{1.2}
\begin{tabular}{|l|c|c|c|c|c|}
\hline
     & All-To-All & Star topology & Square grid  \\ \hline
\textbf{$N^{\textrm{nonint}}_{\textrm{TQG}}$}  & $n-1$         & $n-1$          & $4n-4$    \\ \hline
\textbf{$N^{\textrm{nonint}}_{\textrm{SQG}}$} &  $\frac{5}{2}n+2$         & $\frac{5}{2}n+2$          & $\frac{21}{2}n-\frac{47}{2}$ \\ \hline
\textbf{$N^{\textrm{int}}_{\textrm{TQG}}$}  & $\frac{3}{2}n^2-\frac{3}{2}n$         & $\frac{3}{2}n^2+\frac{3}{2}n-6$          & $3n^2-6n+3$    \\ \hline
\textbf{$N^{\textrm{int}}_{\textrm{SQG}}$}  & $4n^2-\frac{9}{2}n+\frac{7}{2}$         & $4n^2+\frac{7}{2}n-\frac{25}{2}$ & $8n^2-\frac{33}{2}n+\frac{11}{2}$    \\ \hline
\end{tabular}
\caption{Gate count for one Trotter step and for one cycle for different topologies with and without internuclear interactions.}
\label{tab:gate_count}
\end{table}
\end{center}

\section{Qubit-resonator gate theory}
\label{gate_theory}

In the following discussion, we consider gate operation between the resonator and one of the qubits, and neglect any effects that arise from the interactions with spectator qubits and other resonator modes. The time dynamics in such a system are determined by the Hamiltonian: 

\begin{align}
    \label{H2QG}
	\begin{split}
	H= H_0 + H_{rc} + H_{qc} + H_{rq},
	\end{split}
\end{align}
where the uncoupled part  of the total Hamiltonian $H_0 = H_r + H_c + H_q $ is:

\begin{align}
	\begin{split}
	H_r&=\hbar \omega_r b_r^\dagger b_r,\\
	H_c&=\hbar \omega_c b_c^\dagger b_c + \frac{\hbar}{2} \alpha_c b_c^\dagger b_c^\dagger b_c b_c, \\
	H_q&=\hbar \omega_q b_q^\dagger b_q + \frac{\hbar}{2} \alpha_q b_q^\dagger b_q^\dagger b_q b_q,
	\end{split}
\end{align}
where $b_{\lambda}$ and $\omega_{\lambda}$ are the annihilation operator and fundamental frequency for the mode $\lambda=\{r,c,q\}$, respectively, and $\alpha_{\gamma}$ is the anharmonicity of the mode $\gamma=\{q,c\}$. The interaction component of the Hamiltonian is:

\begin{align}
	\begin{split}
H_{\lambda \mu} = -\hbar g_{\lambda \mu}(b^{\dag}_{\lambda}- b_{\lambda})( b^{\dag}_{\mu}- b_{\mu}),
 	\end{split}
 \end{align}
where $\lambda \mu=\{rc,qc,rq\}$, and $g_{\lambda \mu}$ denote resonator-coupler, qubit-coupler and resonator-qubit coupling frequencies. With the Hamiltonian of Eq.~\eqref{H2QG}, we are now in a position to perform simulations of two-qubit gates by propagating a suitably chosen initial state.

Before the gate operation, we choose the idling frequencies for the qubit, resonator, and the coupler such that the CZ coupling rate $\zeta$ is minimized. This CZ coupling rate is defined as:

\begin{align}
\label{ZZdef}
	\begin{split}
	\zeta = \omega_{\textrm{101}} - \omega_{\textrm{100}} - \omega_{\textrm{001}} + \omega_{\textrm{000}},
	\end{split}
\end{align}
where $\omega_{n_r0n_q}$ corresponds to the eigenenergy of Hamiltonian in Eq.~\eqref{H2QG} with $n_r$ excitations in resonator and $n_q$ excitations in qubit with coupler being in the ground state. The point of minimal $|\zeta|$ is also known as the idling configuration, which we found to be at $[\omega_{\rm r},\omega_{\rm c},\omega_{\rm q}]/(2\pi) = [4.30, 6.14, 4.47]$ GHz for the parameters given in Table~\ref{tab:ndonis_parameters}. The CZ gate is operated by sending a flux pulse that modifies the coupler frequency $\omega_c$, which then in the coupled basis modifies the frequencies $\omega_{\textrm{101}},\omega_{\textrm{100}},\omega_{\textrm{001}}$ and $\omega_{\textrm{000}}$. This makes $\zeta$ non-zero, so the system collects a CZ phase. 

\twocolumngrid
\newpage
\bibliography{Qsensing}

\end{document}